\documentclass[prx,10pt,aps,twocolumn,amsmath,amssymb,altaffilletter,superscriptaddress]{revtex4-2}

\usepackage                             {lmodern}
\usepackage     [T1]                    {fontenc}
\usepackage                             {xcolor}
\usepackage                             {graphicx}
\usepackage     [english]               {babel}
\usepackage     [colorlinks,citecolor=blue,linkcolor=blue,urlcolor=blue,bookmarks=false,hypertexnames=true]           {hyperref}
\usepackage                             {dcolumn}
\usepackage                             {bm}
\usepackage                             {textcomp}

\begin{document}

\title{Long-lived coherence in driven many-spin systems: from two to infinite spatial dimensions}

\author{Walter Hahn}
\email{walter.hahn@uibk.ac.at}
\affiliation{QuTech, Delft University of Technology, Lorentzweg 1, 2628 CJ Delft, The Netherlands}
\affiliation{Institute for Quantum Optics and Quantum Information of the Austrian Academy of Sciences, Innsbruck, Austria}

\author{V. V. Dobrovitski}
\email{v.v.dobrovitski@tudelft.nl; author to whom correspondence should be addressed}
\affiliation{QuTech, Delft University of Technology, Lorentzweg 1, 2628 CJ Delft, The Netherlands}
\affiliation{Kavli Institute of Nanoscience, Delft University of Technology, Lorentzweg 1, 2628 CJ Delft, The Netherlands}

\begin{abstract}
Long-lived coherences, emerging under periodic pulse driving in the disordered ensembles of strongly interacting spins, offer immense advantages for future quantum technologies, but the physical origin and the key properties of this phenomenon remain poorly understood. 
We theoretically investigate this effect in ensembles of different dimensionality, and predict existence of the long-lived coherences in all such systems, from two-dimensional to infinite-dimensional (where every spin is coupled to all others with similar strength), which are of particular importance for quantum sensing and quantum information processing. 
We explore the transition from two to infinite dimensions, and show that the long-time coherence dynamics in all dimensionalities is qualitatively similar, although the short-time behavior is drastically different, exhibiting dimensionality-dependent singularity.
Our study establishes the common physical origin of the long-lived coherences in different dimensionalities, and suggests that this effect is a generic feature of the strongly coupled spin systems with positional disorder.
Our results lay out foundation for utilizing the long-lived coherences in a range of application, from quantum sensing with two-dimensional spin ensembles, to quantum information processing with the infinitely-dimensional spin systems in the cavity-QED settings.
\end{abstract}

\maketitle

\section{Introduction}

Collective quantum coherences of many-spins systems play central role in quantum science and technology. But quantum coherence is fragile, and extending its lifetime is a critical problem. For instance, in spin ensembles the collective coherence (collective transverse polarization) is  destroyed by dipolar interactions~\cite{slichter,abragam,haeberlen,ChoCoryEtal}, and the the spin echo signal, which quantifies coherence, quickly decays on the time scale $T_2$~\cite{hahn_echo}. Coherence can be preserved e.g.\ via pulse and/or continuous-wave decoupling that suppresses dipolar interactions~\cite{slichter,abragam,haeberlen}. Recently, an intriguing alternative has attracted much attention: it exploits, rather than fights, the spin-spin interactions. Namely, the unusual many-spin states, which are formed in ensembles of dipolar-coupled spins under periodic driving by $\pi$-pulses, exhibit collective spin coherences living up to $10^5$ times longer than the $T_2^*$ and about $10^4$ time longer than the $T_2$ time~\cite{li,dong,frey,dementyev}. This phenomenon, along with other similar effects, has been observed in various solid-state nuclear magnetic resonance (NMR) experiments~\cite{erofeev,rhim,dementyev,li,dong,Zhang,feldman}, but still remains poorly understood. The long-lived coherences emerge from the combination of strong  dipolar coupling, disorder, and pulse imperfections (understood broadly as deviations of the real $\pi$-pulses from perfect instant 180$^{\circ}$ rotations)~\cite{Zhang,li}. Besides, the long-lived coherences demonstrate subharmonic response, i.e.\ asymmetry in the magnitudes of even and odd echoes~\cite{dementyev,li,Zhang}. Both effects of long coherence lifetime and its subharmonic response are remarkably stable against perturbations and decoherence.

\begin{figure}[]
  \centering
  \includegraphics[width=\columnwidth]{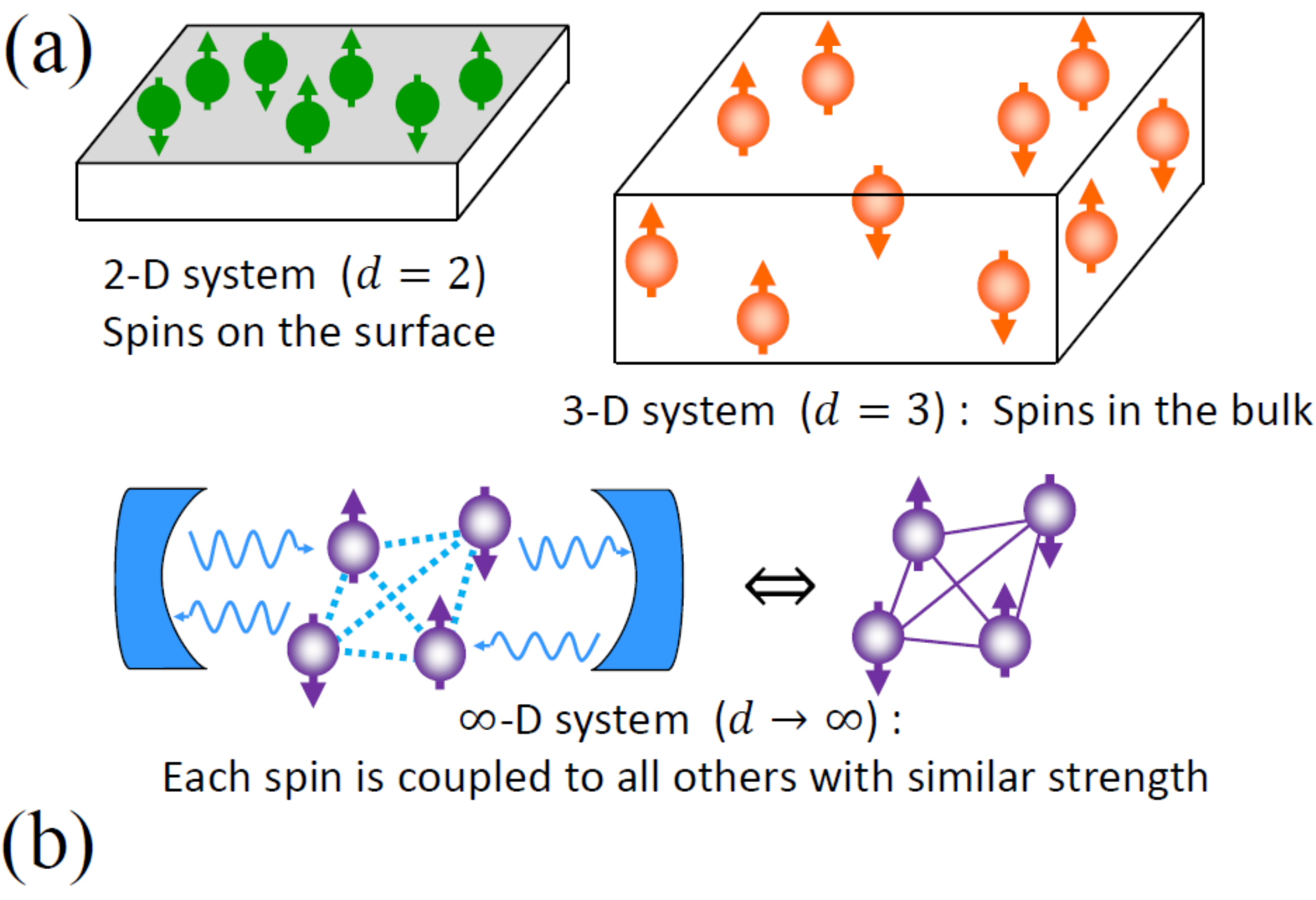}
  \includegraphics[width=\columnwidth]{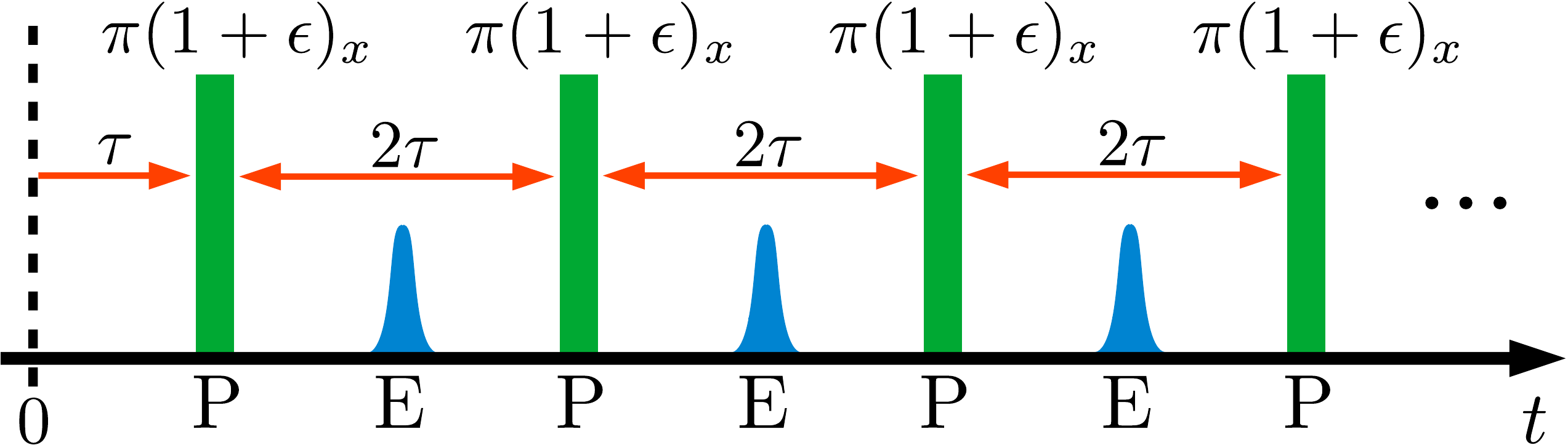}
  \caption{\textbf{(a)} Spin ensembles of different dimensionality: two-dimensional, three-dimensional, and (effectively) infinite-dimensional systems. The latter system with all-to-all interactions of similar strength is realized e.g.\ by coupling the spins to a detuned collective photon/phonon/magnon mode in a cQED-like setting. \textbf{(b)} Schematic representation of the periodic pulse sequence. Imperfect $\pi$ pulses (P) are applied at times $(2n+1)\tau$ ($n\geq0$), so   that a series of echoes (E) is formed at times $2n\tau$. Each pulse rotates the spins along the $x$-axis by an angle $\pi(1+\epsilon)$, where $\epsilon$ is the pulse imperfection parameter.}
 \label{fig_sys}
\end{figure}

The long-lived coherences could be of great benefit for new quantum technology platforms. For instance, promising platforms for quantum sensing utilize two-dimensional systems ($d=2$, see Fig.~\ref{fig_sys}), such as surface spins or 2D layers of NV spins~\cite{wrachtrup,aniajayich,ndeleon,DeWitOosterkampEtal}), 
which can be brought close to the system being sensed, thus improving resolution and sensitivity. 
Employing ensembles of spins boosts the total signal, and thus greatly improves the signal-to-noise ratio, but the collective coherence decays quickly due to dipolar coupling between the spins. Increasing the lifetime of collective coherences would be of enormous benefit for quantum sensing.
On the opposite end ($d\to\infty$) are the spin ensembles in a cavity QED-type settings, actively explored for quantum information applications~\cite{AwschHansonEtalReview,AstnerSchmidmeyerEtal}, where each spin is coupled to all others with a similar strength via collective photonic, phononic, or magnonic mode~\cite{JayichMechDriving,ueda,JayichReview,FuchsMechDrivingHBAR,AndrichAwschalomEtal,DumeigeBudkerEtal,JahnkeLukinJelezko2015}. 
Taking full advantage of the long-lived coherences could increase the signal-to-noise ratio in these systems by orders of magnitude. In order to achieve that,  detailed understanding the long-lived coherences in systems of different spatial dimensionality $d$ is required. So far, even existence of the long-lived coherences at $d=2$ or $d\to\infty$ has remained elusive, and their properties have been unknown. 

In this article we predict that the long-lived coherences do exist in these important systems, thus opening the way to employing them in novel quantum information platforms. In order to analyze in detail the transition from $d=2$ to $d\to\infty$, and to clarify generic features of the long-lived coherence dynamics, we numerically simulate the dynamics of disordered dipolar-coupled quantum spin ensembles of different dimensionalities, subject to periodic driving by imperfect $\pi$-pulses, with the rotation angle slightly deviating from 180$^\circ$. For all dimensionalities $d$ studied, we observe the long-lived coherences and see emerging subharmonic response when the time interval $\tau$ between the pulses increases to become comparable to $T_2$, see Fig.~\ref{fig_sys}b. Our simulations show that the magnitude of the long-lived coherence decreases at larger $d$, but still remains quite large even at $d\to\infty$.
By analyzing the Floquet operator, we establish the kinematic origin of the long-lived coherences, determine that it is similar for all dimensionalities, and identify the states that are involved in their formation. 

Some aspects of the long-lived coherence resemble the time crystal dynamics in periodically driven spin systems~\cite{khemani,yao,yao2,moessner,abanin,abanin2,PolkovnikovBukov2019,Sacha_2017,khemaniNMR}, with their characteristic robustness~\cite{lesanovsky,lazarides,demler}. Time crystals have been observed in many spin systems, from trapped ions~\cite{monroe} to spin ensembles in diamond~\cite{choichoi}, including the kind of NMR systems that exhibits the long-lived coherences~\cite{rovny,sreejith,khemaniNMR}. However, the physical origins of the two phenomena are different: the coherences are determined by the transverse magnetization, while the time crystal dynamics refers to the behavior of the longitudinal polarization. They are governed by different processes, and their lifetimes can differ by many orders of magnitude~\cite{slichter,abragam}; e.g., in the case of perfect pulses (instantaneous 180$^\circ$ rotations) the time crystal oscillations have the longest lifetime and largest amplitude, while the long-lived coherences completely vanish~\cite{Zhang,li,dementyev}. The relation between the long-lived coherence and the time crystal dynamics is poorly explored, in spite of its fundamental interest. Besides, the studies of time crystals so far have mostly focused on $d=1$ \cite{ponte,else,lazarides,keys1,keys2,liu,angelakis,zinner,demler,nunnenkamp,khemaniNMR} and $d=3$ systems~\cite{rovny,sreejith,monroe,choichoi,wenho}, as well as on the systems with $d\to\infty$~\cite{fazio,lesanovsky,nunnenkamp,choichoi,wenho,ueda}.
For these reasons, we also explore here the time crystal dynamics of the longitudinal spin polarization for different dimensionalities. We find that it also demonstrates time crystal-like oscillations with a very long lifetime for all $d$, but the physical features are markedly different from those of the long-lived coherences. More detailed exploration of this issue in the future would be of utmost interest, but is beyond the scope of the present paper.

Dimensionality plays a key role in the dynamics of dipolar-coupled positionally disordered spin systems. Such systems exhibit characteristic $d$-dependent dynamical singularities: the spin dynamics at short times is strongly singular for $d=2$, while for $d\to\infty$ the singularity disappears~\cite{feldman_lacelle,lacelle,dobrovitskiEns,abragam,klauderanderson}. Besides, dimensionality is well known to be decisive in the context of localization and thermalization dynamics in spin ensembles~\cite{andersonLoc,levitov,burin,yao,demler}. Since so many key features of the spin dynamics depend on $d$, one would expect that the properties of long-lived coherences would also strongly depend on $d$. 
Surprisingly, our results demostrate that this expectation is incorrect.


\section{Qualitative discussion of the effect}
\label{sec:qualitative}

We study an ensemble of $N_s$ spins $S_i=1/2$ ($i=1\dots N_s$) in a standard setting of a magnetic resonance-type experiments. Namely, the spin system is placed in a strong quantizing magnetic field $H_Q$ directed along the $z$-axis~\cite{slichter}; this field induces fast spin precession with Larmor frequency $\omega_Q$, which is much larger than all other frequency scales of the problem
\footnote{
In some experimental situations~\cite{choichoi,rovny,wenho} the role of strong  quantizing field is played by strong continuous-wave Rabi driving applied at the frequency $\omega_Q$. In this case, the dynamics is to be considered in a doubly rotating frame, 
where the effective quantization axis is directed along the Rabi driving in the primary rotating frame, and the role of the principal Larmor frequency $\omega_Q$ is played by the Rabi frequency $\omega_R$, see e.g.\ Refs.~\onlinecite{abragam,slichter} for details. The doubly rotating frame in the theory of magnetic resonance is analogous to the dressed state basis in the quantum optics context.
}. 
Following the standard theory of magnetic resonance, we describe spin dynamics in the coordinate frame that rotates around the $z$-axis with the circular frequency $\omega_Q$, and 
retain only the secular terms in the system's Hamiltonian, which remain static in the rotating frame
\footnote{
Note that the conventional rotating-frame description includes {\it two\/} equally important components: (i) the non-secular Hamiltonian terms are usually either dropped or taken into account perturbatively (like the Bloch-Siegert shift of the resonance line), and, (ii) the observables (such as $M_x$ and $M_y$) are also viewed in the rotating coordinate frame, not in the laboratory frame. Correspondingly, all expectation values should be understood as the expectation values in the rotating coordinate frame; in the laboratory frame, even within secular approximation, all graphs and formulas would have a different form, with different dependence on time.
},
or vary slowly in comparison with $\omega_Q$~\cite{slichter,abragam}.

Initially, by applying a preparatory $\pi/2$ pulse, the spins are 
prepared in a state 
weakly polarized along the $x$-axis of the rotating frame, such that the initial ensemble density matrix is $\rho(t=0)\propto \mathbb{I}-\mu M_x$, where $\mathbb{I}$ is identity matrix, $\mu\ll 1$ is a parameter determining the absolute polarization of the ensemble, and $M_x=\sum_j S_{jx}$ is the collective coherence operator. Here and below, $S_{j\alpha}$ with $\alpha=\{x,y,z\}$ denotes the component of the $j$-th spin along the rotating-frame axis $\alpha$. In experiments, the ensemble coherences along the $x$- and $y$-directions are quantified by the total transverse magnetizations $M_x$ and $M_y$ along the corresponding axes, so we use the terms ``coherence'' and ``transverse magnetization'' interchangeably. The longitudinal polarization, exhibiting time crystal-like behavior (see Sec.~\ref{sec:perptodriving}), is quantified by the magnetization $M_z=\sum_j S_{jz}$  along the $z$-axis. Note that in this work we vary only the spatial dimensionality $d$ of the ensemble, while spins themselves remain embedded in three dimensions, i.e.\ have three orthogonal components.

\begin{figure}[]
  \centering
  \includegraphics[width=\columnwidth]{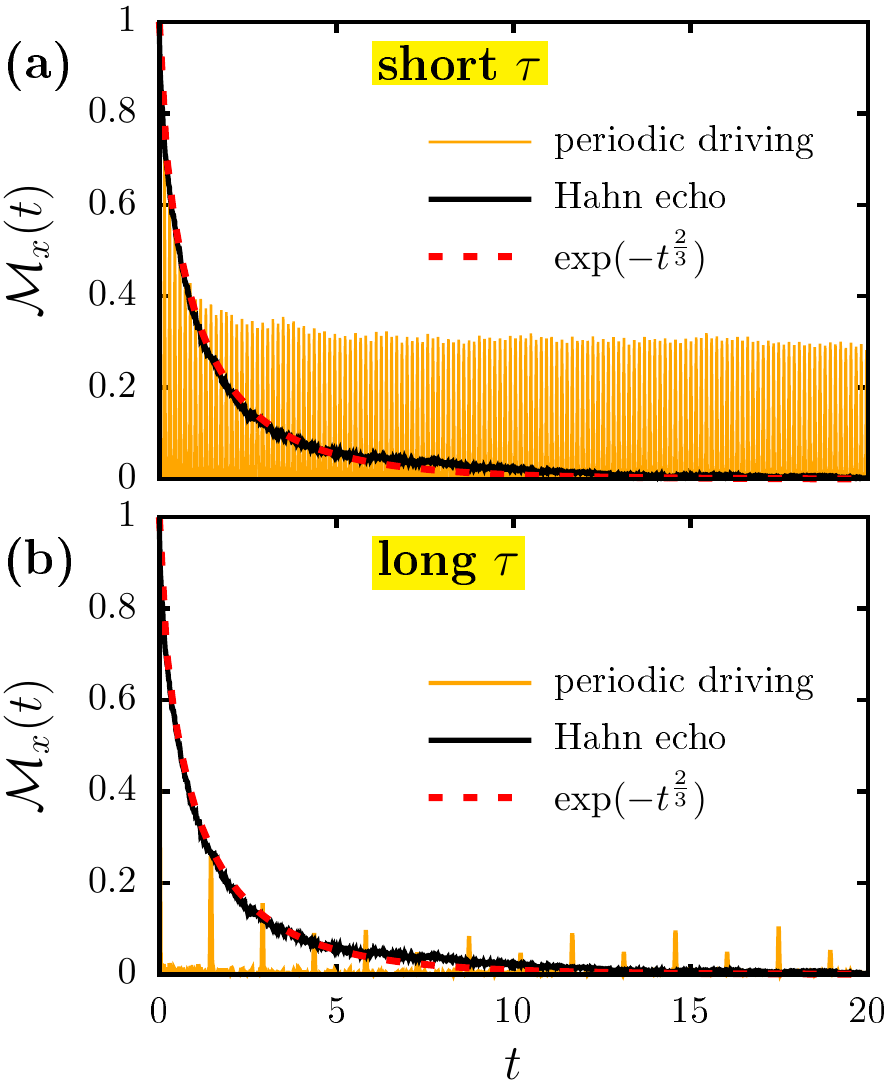}
  \caption{Long-lived coherence $\mathcal{M}_x(t)$ for two-dimensional $d=2$ periodically driven disordered spin systems, for short inter-pulse time delay $2\tau$ ($\tau/T_2\approx0.07$) \textbf{(a)} and for long delay $2\tau$ ($\tau/T_2\approx0.7$) \textbf{(b)}. The long-lived train of echoes $\mathcal{M}_x(t)$ (orange lines) extends far beyond the $T_2=1$ time. The single-pulse Hahn echo (black solid lines) was obtained by simulating a $\pi$ pulse applied at time $\tau=t/2$; the analytical result for the Hahn echo in $d=2$ is $\mathcal{M}_x(t)=\exp(-t^{2/3})$, and is in excellent agreement with numerics. The amplitude of the long-time tails is larger for short $\tau$. For long $\tau$, the even-odd echo asymmetry becomes pronounced. The system size is $N_s=20$, other parameters are $\epsilon=0.07$ and $T_2^*\approx 0.02$. 
  }
 \label{fig_2dtime}
\end{figure}

In typical experiments, the spins experience random quasi-static local magnetic fields, described by the Hamiltonian ${\cal H}_\text{L}=\sum_{j=1}^{N_s}h_jS_{jz}$; everywhere in this article we set $\hbar=1$ and normalize the spins' gyromagnetic ratio to $\gamma=1$. 
\footnote{In this work the local fields $h_j$ are taken as independent Gaussian-distributed random numbers with zero mean and variance $\Gamma^2$, so that the free decay time $T_2^*\sim 1/\Gamma$. The exact form of the distribution is not very important: our simulations with other distributions, such as Lorentzian and exponential, gave essentially the same results. Similarly, the exact values of $\Gamma$ and $T_2^*$ are unimportant: the results remain essentially the same as long as $T_2^*$ is much shorter than all other timescales (except for the width of the driving pulses, which are modeled as instantaneous rotations).} 
These fields cause fast dephasing: the $x$-component of each spin $S_{jx}$ oscillates at its own rate proportional to $h_j$, and the collective coherence $\langle M_x(t)\rangle={\rm Tr}[\rho(t) M_x]$ vanishes at the timescale $T_2^*$. 
This is usually too short for practical needs, and dephasing is suppressed by applying a number of hard $\pi$ pulses, which reverse the sign of the Hamiltonian ${\cal H}_\text{L}$ (ideal, i.e.\ instantaneous 180$^\circ$ hard pulse along the $x$-axis performs rotation $S_{iz}\to -S_{iz}$, $S_{iy}\to -S_{iy}$ for all spins at once). A pulse applied at $t=\tau$ restores collective coherence, producing Hahn echo signal at $t=2\tau$~\cite{slichter,abragam} (see Fig.~\ref{fig_sys}b). 

However, the hard $\pi$ pulses do not affect the spin-spin interaction that    destroys the ensemble coherence by entangling different spins~\cite{slichter,abragam,haeberlen,ChoCoryEtal}.
In relevant experiments, the dominant interaction is the dipolar coupling, described by the Hamiltonian~\cite{slichter,abragam}
\begin{equation} 
\label{eqn_hami}
 {\cal H}_\text{I}=\sum_{i=1}^{N_s} \sum_{j>i} (J_{ij}/2) \left(2S_{iz}S_{jz}-S_{ix}S_{jx}-S_{iy}S_{jy}\right),
\end{equation}
where $J_{ij}=(1-3\cos^2\theta_{ij})/r_{ij}^3$ is the coupling constant between the spins $i$ and $j$, $r_{ij}=|\vec{r}_{ij}|$ is the distance between them, and $\theta_{ij}$ is the polar angle of $\vec{r}_{ij}$. 
Under the influence of ${\cal H}_\text{I}$, the Hahn echo signal gradually decays as a function of time $t=2\tau$ at the timescale $t\sim T_2\gg T_2^*$. 

Note that the dipolar interaction is long-ranged, so that 
each spin is coupled to all other spins 
for all systems considered here, for all values of $d$ from 2 to $\infty$, and the coupling strength $J_{ij}$ decays with distance $r_{ij}$ between the spins $i$ and $j$ in the same way for all $d$. However, the statistical properties of the values $J_{ij}$ greatly vary with the system's dimensionality, thus leading to dramatically different decay of the Hahn echo signal.

The rate of the Hahn echo decay is governed by the positional disorder of spins. In relevant experiments~\cite{dementyev,erofeev,rhim,choichoi,wrachtrup,aniajayich,ndeleon,DeWitOosterkampEtal,AwschHansonEtalReview,AstnerSchmidmeyerEtal}, 
the spins are very dilute: they occupy a small fraction of the lattice sites, and are randomly distributed in the sample. As a result, each spin $S_i$ has its own set of dipolar couplings $J_{ij}$ to the other spins, such that different spins ``feel'' different environments made of other spins. In low-dimensional ensembles these spin-to-spin variations are very strong~\cite{dobrovitskiEns,feldman_lacelle,lacelle}, leading to a very fast decay of the collective coherence. In Appendix~\ref{app_StatAnalysis} 
we show that, in the limit of $T_2\gg T_2^*$, when the flip-flop terms can be omitted and the Hamiltonian (\ref{eqn_hami}) acquires the form $\sum J_{ij} S_{iz}S_{jz}$, the Hahn echo signal $\langle M_x(t)\rangle$ decays with time as
\begin{equation} 
\label{eqn_hahn2}
\langle M_x(t)\rangle\propto\exp\left(-\left|\,t/T_2\right|^{d/3}\right).
\end{equation}
for $d\leq 5$, such that for $d=2$ the initial decay is infinitely fast. For larger $d$ the fluctuations are not as strong, and the singularity of 
$\langle M_x(t)\rangle$ at $t=0$ is weak for $d=4$ and 5. In the limit $d\to\infty$ each spin is coupled to all others with almost uniform coupling, so the echo decay acquires Gaussian form  
$\langle M_x(t)\rangle\propto\exp\left[-\left(t/T_2\right)^2\right]$, without any singularity at all. 

The total Hamiltonian of the system, taking into account the pulse driving, is 
\begin{equation} \label{eqn_htot}
 {\cal H}(t)={\cal H}_\text{I}+{\cal H}_\text{L}+{\cal H}_\text{P}(t),
\end{equation}
where ${\cal H}_\text{P}(t)$ describes periodic driving by a train of (generally imperfect) $\pi$-pulses, as shown in Fig.~\ref{fig_sys}b. If the pulses were ideal, they would suppress the dephasing term ${\cal H}_\text{L}$ (see Appendix~\ref{app_StatAnalysis} for details), and leave the dipolar interaction ${\cal H}_\text{I}$ intact. Thus, the echo signal $\langle M_x(t)\rangle$ would not depend on the number of pulses applied during the time $t$, and the echo decay would follow Eq.~\ref{eqn_hahn2}. 

Our results
show that for non-ideal pulses this is true at short times: the dynamics drastically depends on $d$, with the initial decay rate varying from infinity at $d=2$ to zero at $d\to\infty$. At long times, for spin ensembles of all dimensionalities $d$, the spin dynamics is controlled by accumulation of the pulse imperfections. This process depends on the specific pulse sequence~\cite{dementyev,dong,li}; here we focus on the simple and efficient Carr-Purcell-Meiboom-Gill (CPMG) protocol shown in 
Fig.~\ref{fig_sys}b. 
Accumulation of the pulse errors, combined with the dipolar spin-spin coupling, leads to long-lived tails in the echo signal, extending far past $T_2$ time, in the region where the Hahn echo has already vanished. The magnitude of the long-lived coherence tails is particularly large when the inter-pulse time delay $\tau$ is short compared to $T_2$, as seen in Fig.~\ref{fig_2dtime}a for $d=2$. 
The effect is remarkably robust, and does not disappear even when $\tau$ becomes comparable to $T_2$. In that regime, another feature emerges for all values of $d$: the long-lived coherences exhibit pronounced subharmonic response (Fig.~\ref{fig_2dtime}b), with even echoes being larger than odd ones. This behavior reflects the fact that the period of the evolution operator for CPMG protocol is twice the period of the Hamiltonian ${\cal H}_\text{P}(t)$ of the CPMG pulse train~\cite{haeberlen} (see Sec.~\ref{sec_3d}). 
%

We use direct numerical solution of the many-spin time-dependent Schr{\"o}dinger equation with the second-order Suzuki-Trotter~\cite{SuzukiTrotterDecomp} or Chebyshev~\cite{refCheb} expansion of the evolution operator $U(t)$ for up to $N_s=24$ spins. In all situations that we tested, both methods give consistent results. The initial mixed-state density matrix is represented by a random pure state, i.e.\ as a random unit-norm complex-valued vector of length $2^{N_s}$, uniformly sampled from a sphere $S^{2^{N_s}-1}$ of unit radius. We calculate the experimentally measured  normalized magnetization response
\begin{equation} \label{eqn_normag}
 \mathcal{M}_\alpha(t)=\langle M_\alpha(t)\rangle / \langle M_\alpha(0)\rangle,
\end{equation}
where $\alpha=\{x,y,z\}$ and 
$\langle M_\alpha(t)\rangle={\rm Tr}[M_\alpha U(t) M_\alpha U^\dag(t)]$ for initial polarization along the axis $\alpha$. The spins are randomly placed in a $d$-dimensional cube, and averaging is performed over 100--360 realizations of the spatial arrangements and values of local fields. The cube's edge length (i.e.\ the spin density $f_s$) is adjusted to have $T_2=1$ for each system after averaging (see Appendix~\ref{app_StatAnalysis} for relation between $T_2$ and $f_s$). $T_2$ is defined as the time when the echo magnitude decreases by the factor $e$. 
In experiments the pulse imperfections may arise from deliberate or accidental miscalibration, or also due to 
local fields and dipolar interactions, which affect the spin rotation during the finite-duration pulses~\cite{Zhang,li,dementyev}. For hard short pulses, imperfections are small, and in this article we model nonideal pulses as instantaneous rotations around the $x$-axis by an angle $\pi(1+\epsilon)$ with $\epsilon\ll 1$.

\begin{figure}[]
  \centering
  \includegraphics[width=\columnwidth]{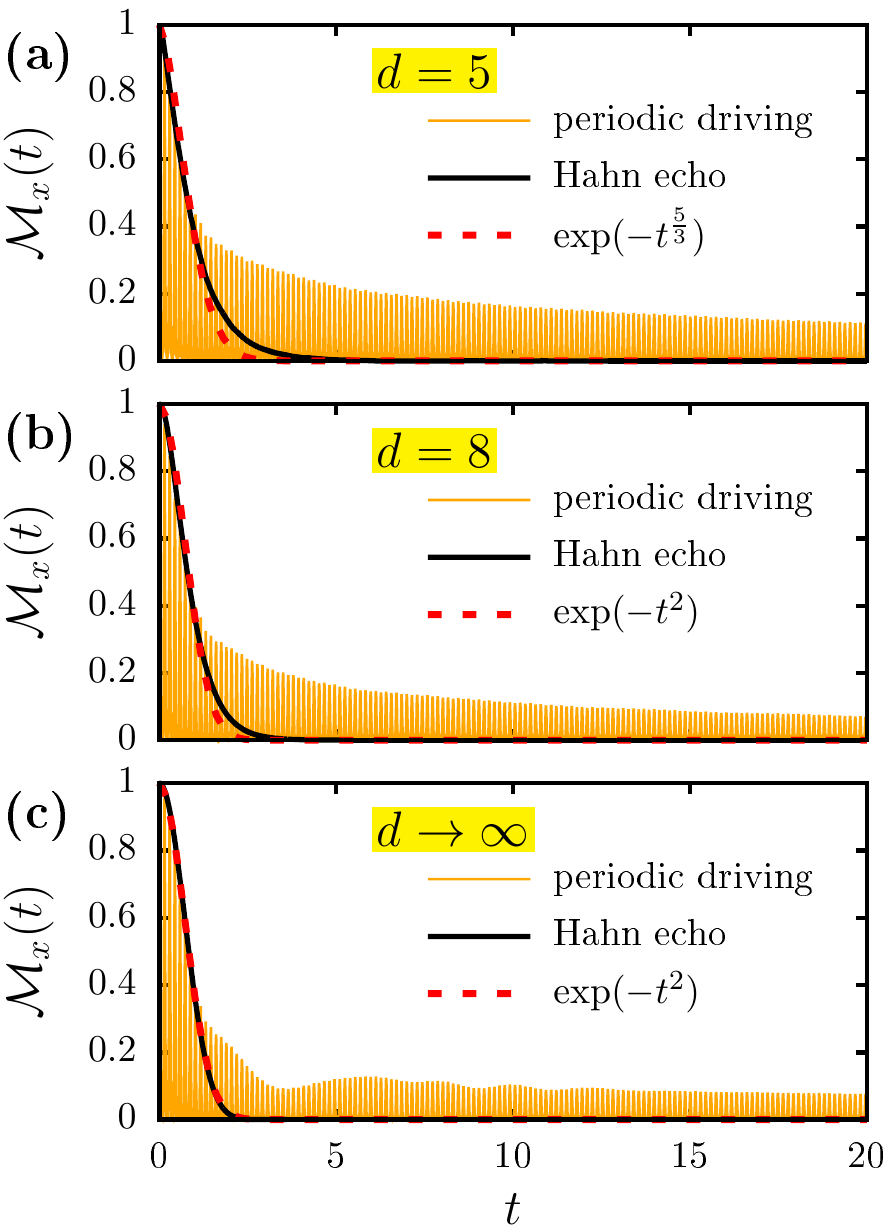}
  \caption{Long-lived coherence $\mathcal{M}_x(t)$ for periodically driven disordered spin systems of different spatial dimensionalities $d$,  indicated in the figure. The echoes $\mathcal{M}_x(t)$ extend far beyond the $T_2$ time. For the single-pulse Hahn echo, the $\pi$-pulse is applied at time $\tau=t/2$. with increasing $d$, the Hahn echo approaches Gaussian form.
  The simulation parameters are $N_s=20$, short $\tau\approx 0.07$, $\epsilon=0.07$ and $T_2^*\approx 0.02$. 
  }
 \label{fig_Idtime}
\end{figure}

\section{Dynamics of the long-lived coherences}
\label{sec_3d}

The behavior of $\mathcal{M}_x(t)$ in pulse driven spin ensembles, as described above, is shown in Fig.~\ref{fig_2dtime} for $d=2$, and in Fig.~\ref{fig_Idtime} for larger $d$. For reference, we presented the results for previously unexplored cases of $d=2$, 5, 8, and $d\to\infty$. For other dimensionalities $d$ that we studied, all results remain qualitatively the same. 
\footnote{
For $d=2$ the spins are located on a plane, and the $z$-axis (direction of the quantizing field) can be aligned at different angles with respect to this plane. If the $z$-axis is normal to the plane, all spin pairs have $\theta_{ij}=\pi/2$, while for the in-plane alignment this angle varies between 0 and $\pi$. However, the resulting changes in the Hahn echo are very small: for a fixed areal spin density $f_s$, the values of $T_2$ differ by only 4\%. The long-lived coherences also do not exhibit any qualitative differences between these two cases. Fig.~\ref{fig_2dtime} presents the results for the case of $z$-axis normal to the plane.}
\footnote{
Due to the nature of the case $d\to\infty$, the averaging for this ensemble is performed only over random local fields $h_j$, while the results for finite $d$ also include averaging over the disorder in spin positions.}


We consider two different values of the inter-pulse delay $\tau$: short $\tau\approx0.07\, T_2$, and long $\tau\approx0.7\, T_2$ (recall that $T_2=1$).
At short times $t\lesssim T_2$, the system's response $\mathcal{M}_x(t)$ closely follows the Hahn echo, and is in excellent agreement with our analytical predictions (see Appendix~\ref{app_StatAnalysis} for details), as seen in the figures (the small differences between the simulated Hahn echo decay and the analytical predictions are due to the finite-size effects). 
%
At longer times $t\gg T_2$, the magnetization response $\mathcal{M}_x(t)$ exhibits long-time tails for all dimensionalities $d$ considered, and for both short and long $\tau$. These long-time tails extend far beyond the $T_2$ time, and in experiments are likely to be limited by the spin-lattice relaxation time $T_1$. With increasing $d$, the overall amplitude of the long-time echoes becomes somewhat smaller. Still, even in the limit $d\to\infty$, the long-time tails do not vanish but saturate at a nonzero value. Also, the amplitude of the long-time tails becomes smaller as the inter-pulse delay $\tau$ increases. 

When the inter-pulse delay $\tau$ is large, comparable to $T_2$, the long-lived coherences demonstrate pronounced subharmonic dynamics (Fig.~\ref{fig_2dtime}b), where the period of the magnetization response $4\tau$ is twice longer than the period of driving $2\tau$. The subharmonic response was observed for all values of $d$ we modeled. This feature can be rationalized with the notion of the cycling period of a pulse sequence~\cite{haeberlen}: for the control Hamiltonian 
${\cal H}_\text{P}(t)$, the cycling period $t_c$ is defined by the condition of periodicity of the evolution operator, i.e.\ $U_1(t)=U_1(t+t_c)$, where $U_1(t)=T\exp[-i\int_0^t{\cal H}_\text{P}(t')dt']$. It is known that the cycling period of a pulse sequence can differ from the period of the Hamiltonian~\cite{haeberlen}, e.g.\ for ideal $\pi$-pulses, the cycling period of the CPMG protocol is $t_c=4\tau$, and includes two $\pi$-pulses~\cite{clock}, i.e.\ contains two periods of the underlying control Hamiltonian ${\cal H}_\text{P}(t)$. We also note that the even-odd asymmetry is present for short $\tau$, but its amplitude is too small to be easily seen.

It is important to point out that the long-lived coherences and the subharmonic response  crucially depend on the driving sequence. For instance, for an alternating-phase Carr-Purcell (APCP) sequence, where the direction of driving alternates between the positive and the negative $x$-direction, no long-lived tails of $\mathcal{M}_x(t)$ emerge~\cite{li}.

\section{Floquet operator analysis of the long-lived coherences}
\label{sec:Floquet}

\begin{figure}[]
  \centering
  \includegraphics[width=\columnwidth]{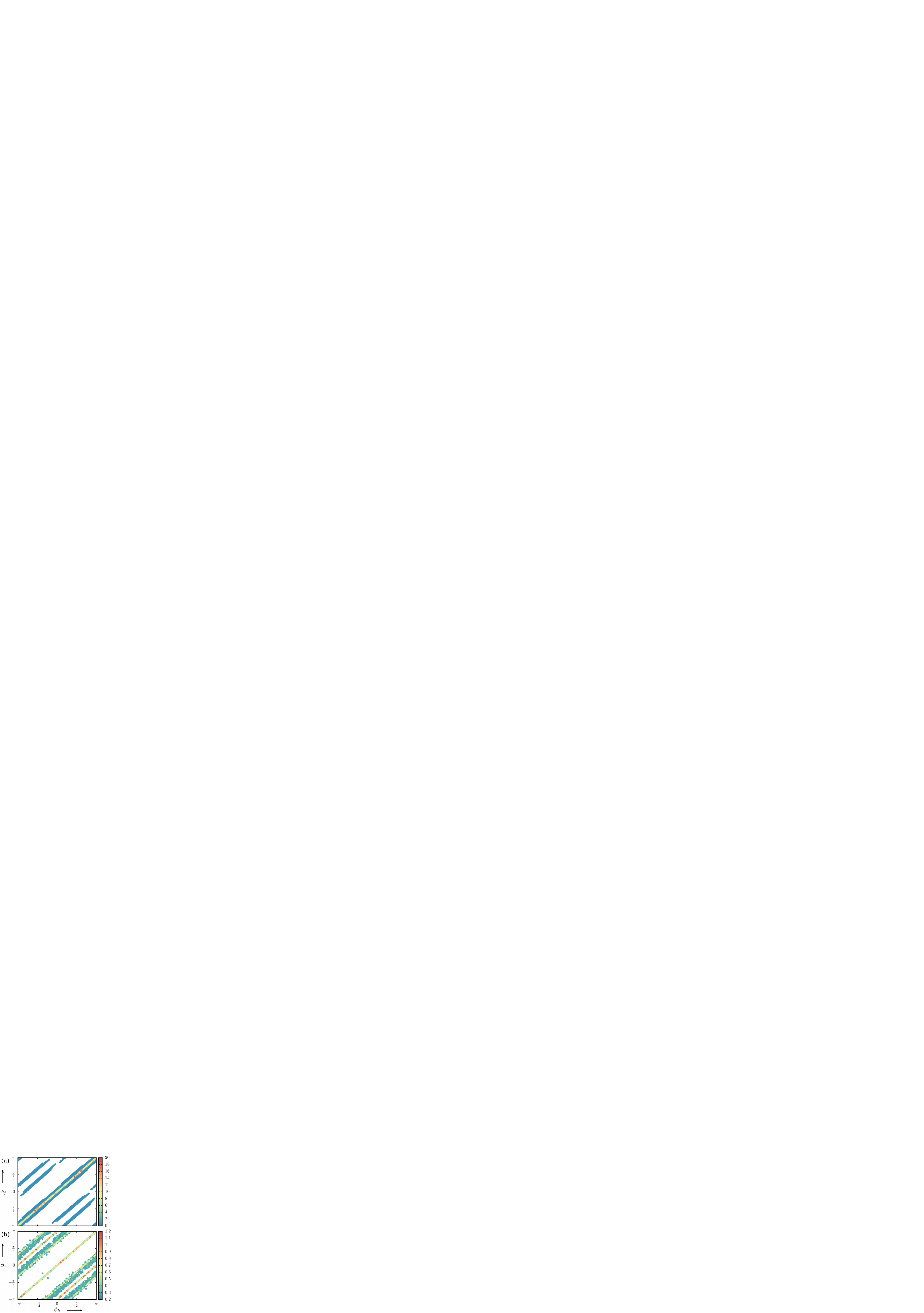}
\caption{The matrix $M^{jk}_{x}=|\langle\psi_j|M_x|\psi_k\rangle|^2$ for short $\tau$ \textbf{(a)} and long $\tau$ \textbf{(b)} for a two-dimensional $d=2$ system. On the vertical and the horizontal axes, respectively, the quasienergies $\phi_j$ and $\phi_k$ of the corresponding Floquet eigenstates $|\psi_j\rangle$ and $|\psi_k\rangle$ are plotted. For short $\tau$ (panel \textbf{a}), a large number of large entries concentrate on the diagonal $\phi_j\approx\phi_k$, producing long-lived coherence with noticeable amplitude. For long $\tau$ (panel \textbf{b}), the values at semi-diagonals $|\phi_j-\phi_k|\approx \pi$ become comparable to the values on the diagonal, which leads to noticeable subharmonic response. Overall, the values for long $\tau$ are smaller than those for short $\tau$. The system size is $N_s=14$. Only values larger than 0.25 are included in the figures.
}
\label{fig_2d}
\end{figure}

Periodically driven systems, described by a Hamiltonian obeying ${\cal H}(t+2\tau)={\cal H}(t)$, can be analyzed using Floquet theory. The stroboscopic time evolution of the system's density matrix, considered only at times that are integer multiples $m$ of the driving period $2\tau$, can formally be written as $\rho(2\tau m)=U_\text{F}^m\rho(0)U_\text{F}^{\dagger m}$, where $U_\text{F}$ is time evolution operator per one period of the driving Hamiltonian (Floquet operator). For the CPMG pulse sequence considered here, the Floquet operator has the form $U_\text{F}=U_{\cal H}(\tau)R_x[\pi(1+\epsilon)]U_{\cal H}(\tau)$, where $R_x[\pi(1+\epsilon)]$ is the operator of rotation around the $x$-axis by an angle $\pi(1+\epsilon)$, and $U_{\cal H}(\tau)$ is the operator of evolution under the action of the system's internal Hamiltonian ${\cal H}_\text{I}+{\cal H}_\text{L}$.

\begin{figure}[] 
  \centering
  \includegraphics[width=\columnwidth]{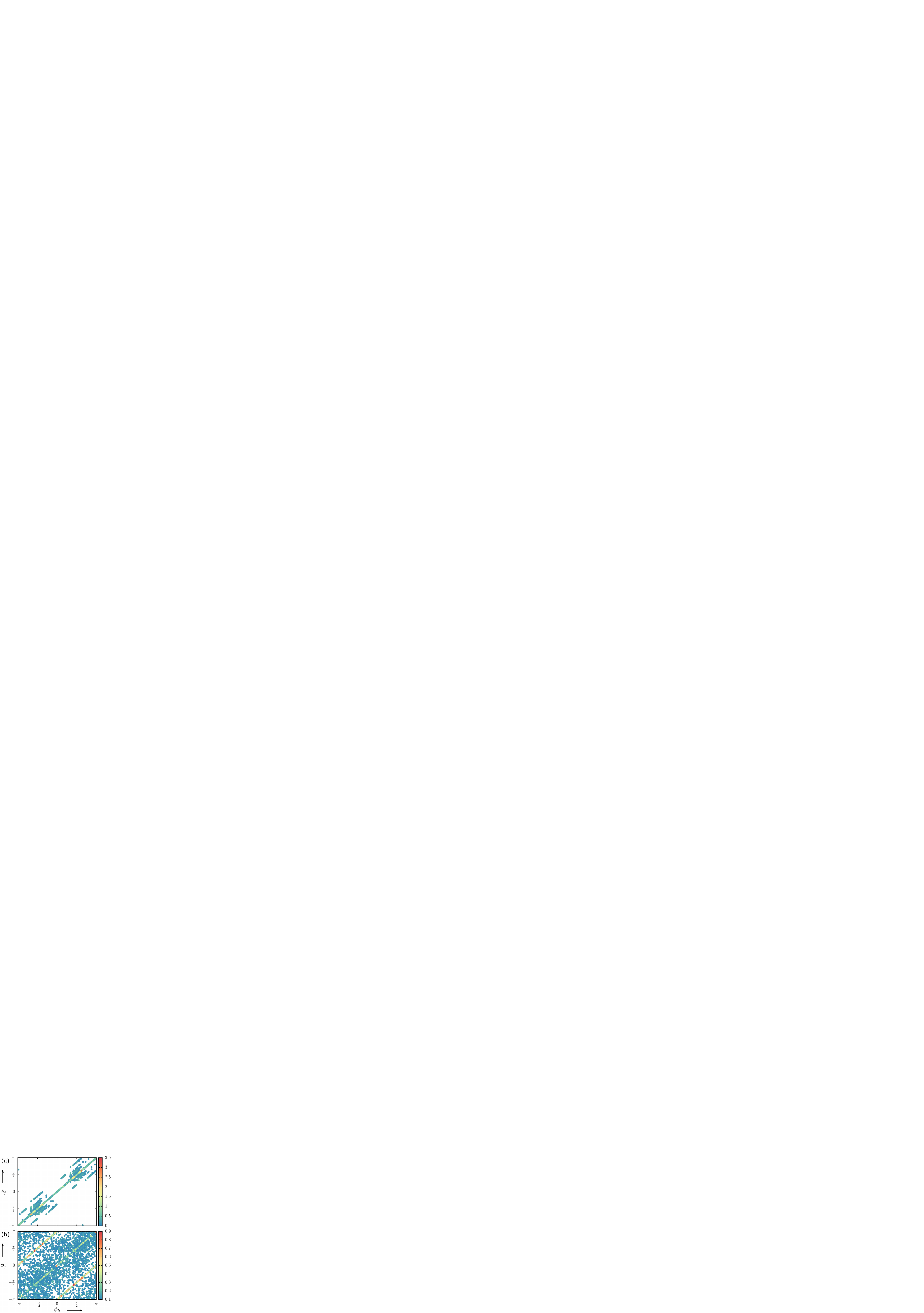}
\caption{Same as Fig.~\ref{fig_2d} but for a five-dimensional $d=5$ system. Only values larger than 0.1 are included in the figures. The same features as in the case $d=2$ are seen along the diagonal $\phi_j\approx\phi_k$ and at the semi-diagonals $|\phi_j-\phi_k|\approx \pi$; these features correspond to the long-lived coherences and to the subharmonic response, respectively. Similarity in the structure of the matrices $M^{jk}_{x}$ for ensembles of different dimensionalities confirms similarity in the physical origin and behavior of the long-lived coherences for different $d$.} 
\label{fig_5d}
\end{figure}

As a unitary operator, $U_\text{F}$ possesses a complete orthonormal set of eigenstates $|\psi_k\rangle$, with complex eigenvalues $e^{i\phi_k}$ of unit modulus:
\begin{equation}
 U_\text{F}=\sum_k e^{i\phi_k} |\psi_k\rangle\langle\psi_k|,
\end{equation}
so the magnetization response after $m$ driving periods is
\begin{equation} \label{eqn_px}
 \mathcal{M}_x(2\tau m) = \frac{4}{N_s2^{N_s}}\sum_{j,k}e^{i(\phi_j-\phi_k)m}|\langle\psi_j|M_x|\psi_k\rangle|^2.
\end{equation}
The signal $\mathcal{M}_x(2\tau m)$ is therefore mainly determined by two quantities: firstly, by the magnitude of the matrix elements
$M^{jk}_{x}=|\langle\psi_j|M_x|\psi_k\rangle|^2$, and, secondly, by the 
distribution $P(\phi_j-\phi_k)$ of the quasienergy differences $\phi_j-\phi_k$, i.e.\ by the number of Floquet eigenstates $|\psi_j\rangle$ and $|\psi_k\rangle$ with a given difference in the quasienergies $\phi_j$ and $\phi_k$. 
The terms with $j=k$ in Eq.~\eqref{eqn_px} do not depend on $m$, so the long-time response $\mathcal{M}_x(2\tau m)$ is governed by the diagonal elements 
$M^{jj}_{x}$ of the matrix $M^{jk}_{x}$. The subharmonic response (the even-odd echo asymmetry) is controlled by the pairs of Floquet eigenstates obeying $\phi_j-\phi_k\approx\pm\pi$, such that $e^{i(\phi_j-\phi_k)m}\approx (-1)^m$.

\begin{figure}[]
  \centering
  \includegraphics[width=\columnwidth]{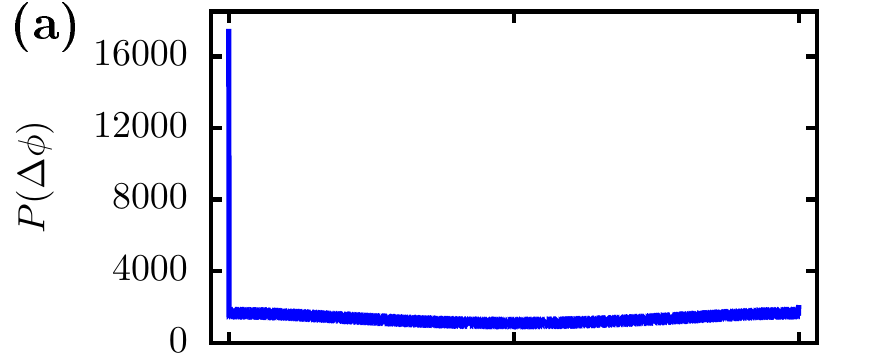}
  \includegraphics[width=\columnwidth]{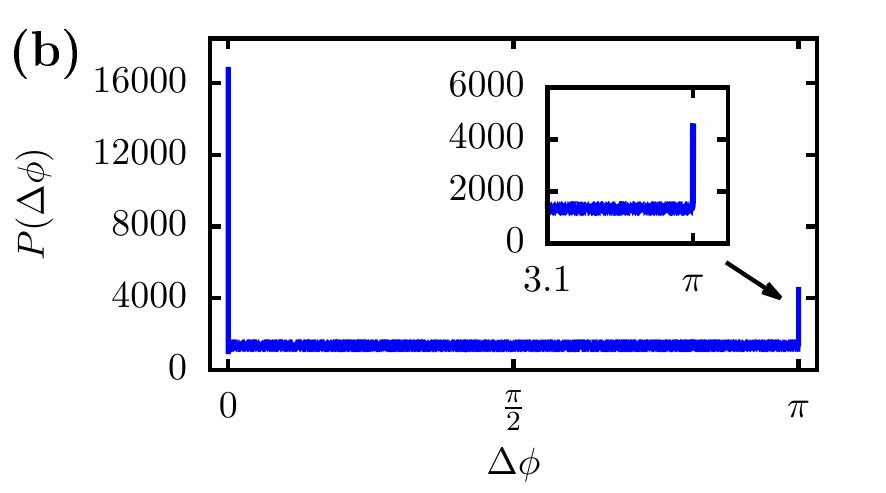}
  \caption{Histogram of the quasienergy differences $\Delta\phi$ for a $d=2$ disordered spin system with $N_s=14$, at short $\tau$ \textbf{(a)} and long $\tau$ \textbf{(b)}. The sharp peaks at $\Delta\phi=0$ and at $\Delta\phi=\pi$ correspond to the long-lived coherences and to the subharmonic response, respectively. Specifically, $P(\Delta\phi)$ is the total number of the Floquet eigenstates $|\psi_j\rangle$ and $|\psi_k\rangle$ having a given difference $\Delta\phi$ in their quasienergies $\phi_j$ and $\phi_k$. The quantity $\Delta\phi$ is downfolded to the interval $\Delta\phi\in [0,\pi]$ and binned, so that $P(\Delta\phi)$ includes all states whose quasienergies satisfy the condition $|\phi_j-\phi_k|\in\left[\Delta\phi-\beta,\Delta\phi+\beta\right]$ or $2\pi-|\phi_j-\phi_k|\in\left[\Delta\phi-\beta,\Delta\phi+\beta\right]$, where $2\beta=\pi\cdot 10^{-5}$ is the width of a bin. To avoid double counting, only the states with $\phi_j\ge\phi_k$ are included in $P(\Delta\phi)$. The results shown are averaged over many realizations of the disorder.
%
%
  The inset in panel \textbf{(b)} shows the magnified view of the peak at $\Delta\phi\approx\pi$. The simulation parameters are the same as in Fig.~\ref{fig_2dtime}.
 }
 \label{fig_specdiff}
\end{figure}

An example of the matrix $M^{jk}_{x}=|\langle\psi_j|M_x|\psi_k\rangle|^2$ is shown in Fig.~\ref{fig_2d} for short $\tau$ (a) and long $\tau$ (b) for a two-dimensional system, for one typical realization of the positional disorder. 
The matrix $M^{jk}_{x}$ for short $\tau$ is dominated by large diagonal elements, whereas the off-diagonal elements are almost negligible. This corresponds to the pronounced long-lived coherences in Fig.~\ref{fig_2dtime}a and Fig.~\ref{fig_Idtime}, with the almost time-independent amplitude. It is clearly seen that the long-lived coherence contains comparable contributions from a large number of Floquet eigenstates, rather than being confined to a small subset of some special states.

For long $\tau$, the matrix $M^{jk}_{x}$ exhibits large entries both on the diagonal, and on the two semi-diagonal lines corresponding to  $\phi_j-\phi_k\approx\pm\pi$; the latter correspond to emerging even-odd echo asymmetry seen in Fig.~\ref{fig_2dtime}b. The semi-diagonals also show comparable contributions from a large number of pairs of Floquet eigenstates. In comparison with the case of short $\tau$, the diagonal values $M^{jj}_{x}$ on  average are smaller for long $\tau$, corresponding to smaller amplitude of the long-time tails in Fig.~\ref{fig_2dtime}b.

The matrices $M^{jk}_{x}$ for other $d$ exhibit similar structure. As an example, Fig.~\ref{fig_5d} exhibits the results for $M^{jk}_{x}$ in the case of $d=5$. The results of diagonalization of the Floquet operator for different $d$ evidence that the physical origin of the effect of the long-lived coherences is similar for all dimensionalities.

The other quantity determining the signal $\mathcal{M}_x(2\tau m)$ in Eq.~\eqref{eqn_px} is the distribution $P(\Delta\phi)$ of the quasienergy differences $|\phi_j-\phi_k|$. In order to take into account the symmetries of  the summands in Eq.~\ref{eqn_px}, we downfold the quantity $\Delta\phi$ to the interval $[0,\pi]$, i.e.\ we take  
$\Delta\phi=|\phi_j-\phi_k|$ when $|\phi_j-\phi_k|\le\pi$, and $\Delta\phi=2\pi-|\phi_j-\phi_k|$ when $|\phi_j-\phi_k|>\pi$ (so that $\Delta\phi$ is the smallest angular distance between $\phi_j$ and $\phi_k$ on the $S^1$ circle).
The binned distribution $P(\Delta\phi)$ for $d=2$ is shown in Fig.~\ref{fig_specdiff}, the bin width is $2\beta=\pi\cdot 10^{-5}$. A peak at $\Delta\phi\approx\pi$ clearly emerges for long $\tau$. This means that the number of quasienergy pairs with $|\phi_j-\phi_k|\approx\pi$ becomes larger for long $\tau$. Hence, the subharmonic response emerges not only because the values of $M^{jk}_{x}$ on the semi-diagonals become larger, but also because the total number of non-zero entries on the semi-diagonals increases.


\section{Longitudinal magnetization and the infinite-temperature time crystal dynamics}
\label{sec:perptodriving}

\begin{figure}[]
  \centering
\includegraphics[width=\columnwidth]{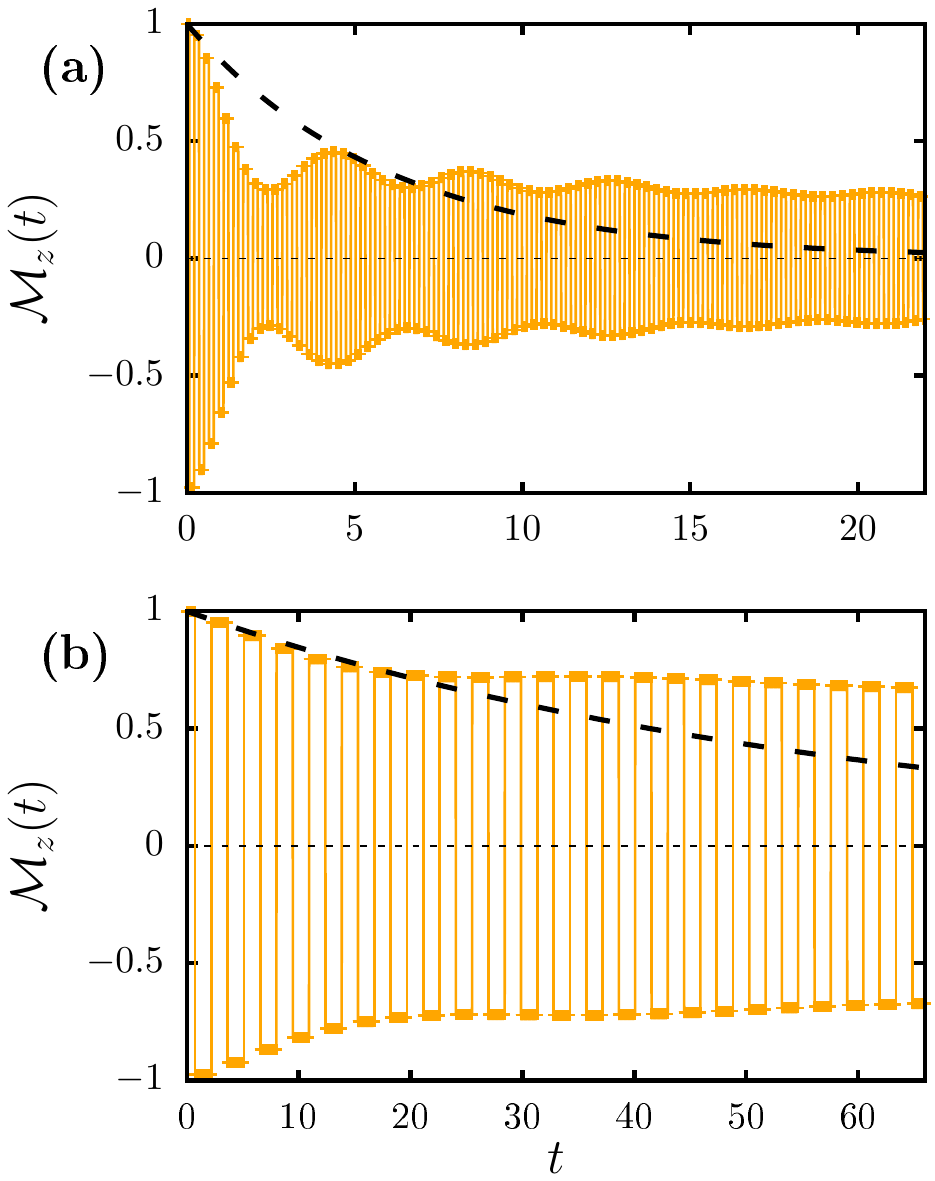}
  \caption{Long-time tails of the $z$-component of collective magnetization $\mathcal{M}_z(t)$ in a $d=2$ disordered spin system for short $\tau$ \textbf{(a)} and long $\tau$ \textbf{(b)} (orange lines and symbols), suggesting time crystal-like behavior. Between the pulses, the value of $\mathcal{M}_z(t)$ is constant because the internal Hamiltonian ${\cal H}_\text{I}+{\cal H}_\text{L}$ conserves the $z$-component of the total magnetization. Each $\pi$-pulse flips the $z$-magnetization, so that the subharmonic response with the period $4\tau$ is the dominating component of the system's long-time response. Without dipolar coupling, accumulated pulse error would modulate the magnetization along the $z$-axis, and $\mathcal{M}_z(t)$ would decay as $\cos^m(\pi\epsilon)$ after $m$ pulses (dashed black line). However, in the presence of the dipolar coupling, $\mathcal{M}_z(t)$ exhibits long-time tails, alternating between the directions ``up'' and ``down'' after each $\pi$-pulse. The system size is $N_s=20$, $\epsilon= 0.07$, and $T_2^*\approx 0.02$.
 }
 \label{fig_2dz}
\end{figure}

Let us now 
focus on 
the dynamics of the longitudinal polarization $\mathcal{M}_z(t)$, which under some circumstances can exhibit robust long-lived infinite-temperature time crystal-like dynamics \cite{khemani}. 

\begin{figure}[]
  \centering
  \includegraphics[width=\columnwidth]{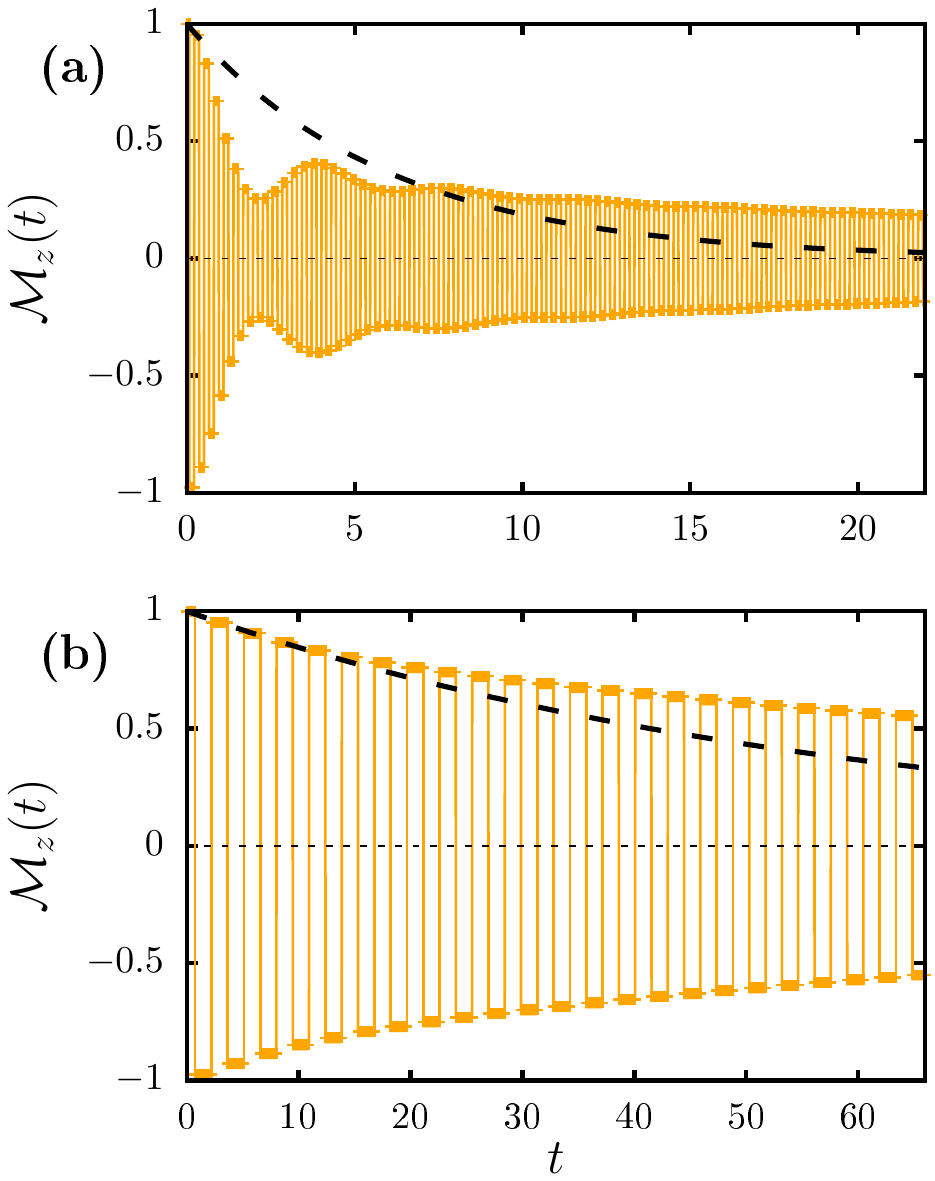}
  \caption{Same as Fig.~\ref{fig_2dz} but for an eight-dimensional $d=8$ disordered spin system.}
 \label{fig_8dz}
\end{figure}

Long-lived coherences and time crystal dynamics share a number of similarities: both are induced by strong spin-spin interactions, robust to experimental imperfections, and demonstrate subharmonic dynamics under appropriate conditions. However, the underlying physics is totally different. 
Since the system's internal Hamiltonian ${\cal H}_\text{I}+{\cal H}_\text{L}$ conserves the total $z$-magnetization, the signal $\mathcal{M}_z(t)$ remains constant between the pulses. If the $\pi$-pulses were ideal, then $\mathcal{M}_z(t)$ would just switch between $+1$ and $-1$. In contrast, the coherence $\mathcal{M}_x(t)$ would quickly vanish under the action of the dipolar spin-spin coupling, along with the Hahn echo, at the timescale $T_2$, without any long-lived tail. 

For non-ideal pulses, one would expect the longitudinal polarization to decay with increasing the number of applied pulses. Indeed, for $\epsilon>0$, after an imperfect pulse the absolute value of $\mathcal{M}_z(t)$ would decrease by a factor $\cos(\pi\epsilon)$, while the $y$-component increases by  $\sin(\pi\epsilon)$. Accordingly, if the $y$-component has irreversibly dephased to zero during the time $2\tau$ between pulses, then the absolute value of $\mathcal{M}_z(t)$ would be expected to decay monotonically as $\cos^m(\pi\epsilon)$ with increasing the number $m$ of applied pulses \cite{rovny,khemaniNMR}. 

The actual behavior of the longitudinal magnetization response $\mathcal{M}_z(t)$  in a $d=2$ disordered spin system with $N_s=20$ is shown in Fig.~\ref{fig_2dz}. Initial decay roughly follows the expected $\cos^m(\pi\epsilon)$ pattern for both short and long $\tau$ (panels (a) and (b), respectively); the additional modulation seen in panel (a) is likely due to incomplete dephasing of the $y$-component between the pulses due to short $\tau$. 

At later times, however, the absolute magnitude of $\mathcal{M}_z(t)$ stabilizes, showing little (if any) decay, and $\mathcal{M}_z(t)$ itself just alternates between positive and negative values. The period of $\mathcal{M}_z(t)$ response is $4\tau$, which is twice the period of the of the driving Hamiltonian ${\cal H}_\text{P}(t)$, and equals, as expected, the cycling period of the CPMG driving sequence. This long-lived subharmonic response is seen for all dimensionalities we studied: the results for another example, a $d=8$ spin ensemble, are presented in Fig.~\ref{fig_8dz}, and are very similar, except that  the amplitude of the long-time tail is somewhat smaller than in the $d=2$ case.
This conclusion is also consistent with the experimental evidence reported for $d=3$ spin systems~\cite{choichoi,rovny,sreejith}. 

Note that the initial decay of $\mathcal{M}_z(t)$ does not exhibit the singularities present in the short-time dynamics of $\mathcal{M}_x(t)$, because it is governed by a  different physical process, by accumulation of the rotation errors. Likewise, the long-time behavior of the longitudinal $\mathcal{M}_z(t)$ and the transversal $\mathcal{M}_x(t)$ polarization is also qualitatively different. For instance, with increasing the inter-pulse delay $\tau$, the amplitude of the long-time tails of $\mathcal{M}_z(t)$ increases, while the amplitude of $\mathcal{M}_x(t)$ decreases, thus clearly demonstrating the difference between the time crystal dynamics and long-lived coherences. 

\begin{figure}[]
  \centering
  \includegraphics[width=\columnwidth]{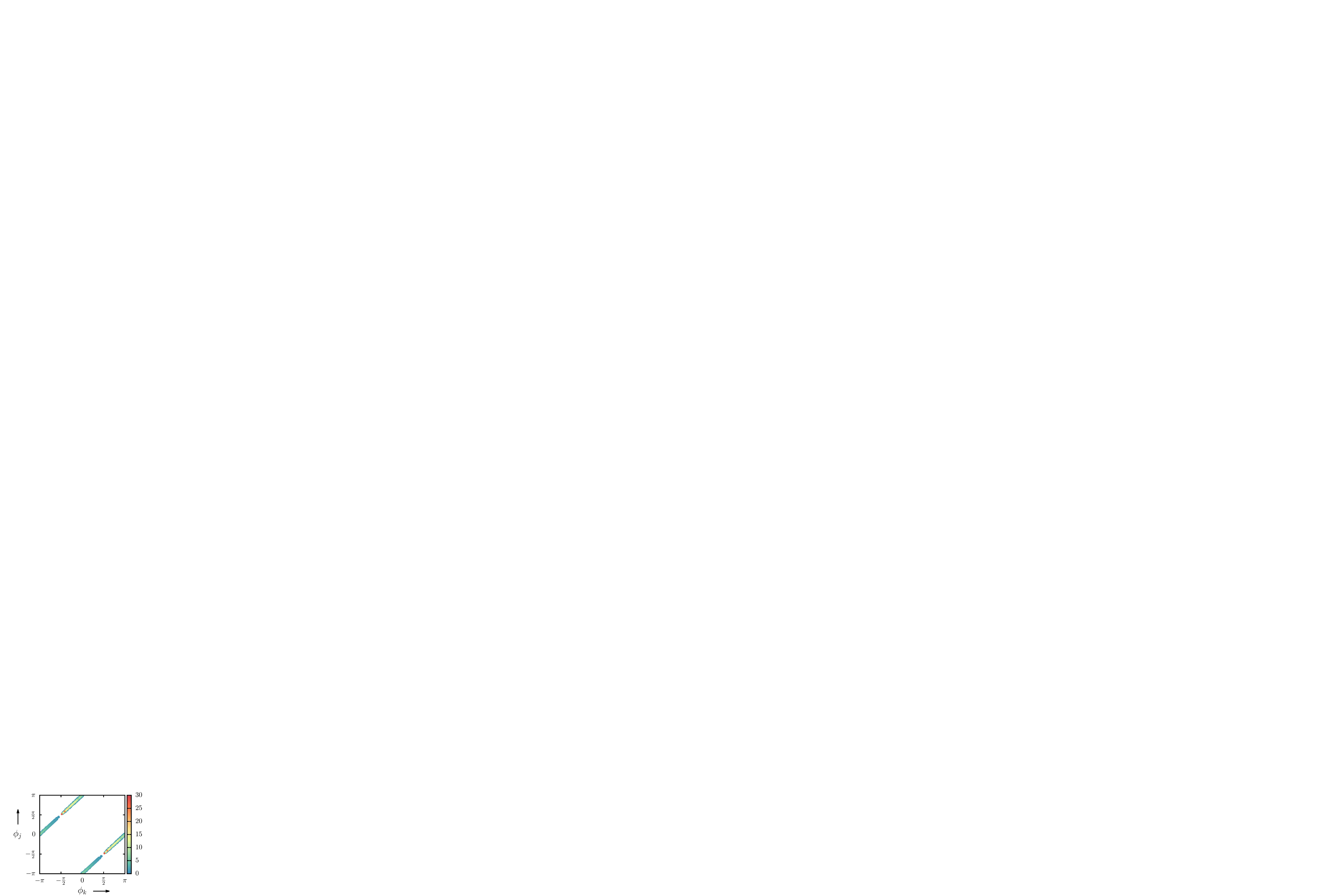}
\caption{A typical example of the matrix $M^{jk}_{z}=|\langle\psi_j|M_z|\psi_k\rangle|^2$ for a two-dimensional $d=2$ system and short $\tau$. The vertical and horizontal axes correspond to the quasienergies $\phi_j$ and $\phi_k$ of the corresponding Floquet eigenstates $|\psi_j\rangle$ and $|\psi_k\rangle$. The entries are concentrated along the semi-diagonals $\phi_j-\phi_k\approx\pm\pi$, evidencing time crystal-like dynamics with the doubled period $4\tau$. For long $\tau$, the structure of the matrix $M^{jk}_{z}$ is very similar (not shown). Only values larger than 1.0 are included in the figure. The system size is $N_s=15$.}
\label{fig_2dzfl}
\end{figure}

We have also analyzed the stroboscopic evolution of $\mathcal{M}_z(t)$ by diagonalizing the Floquet operator; an example of the corresponding matrix $M^{jk}_{z}=|\langle\psi_j|M_z|\psi_k\rangle|^2$ is shown in Fig.~\ref{fig_2dzfl} for one typical realization of the positional disorder. As anticipated, it exhibits nonzero values only at the semi-diagonals $\phi_j-\phi_k\approx\pm\pi$, in accordance with the subharmonic response described above. The diagonal elements are negligible. Similar to the case of long-lived coherences, we see again that many states, distributed all over the Hilbert space, contribute to the effect. 

For the initial polarization along the $y$-axis, no long-lived magnetization response $\mathcal{M}_y(t)$ is seen in any dimensionality $d$, and the elements of the matrix $M^{jk}_{y}=|\langle\psi_j|M_y|\psi_k\rangle|^2$ are also generally small, without any clear structure.

\section{Final remarks and conclusions} 
\label{sec_conc}

Before concluding, let us make some final remarks.

a) Time crystal dynamics and the long-lived coherences exhibit striking similarity: both are induced by spin-spin interactions and demonstrate subharmonic dynamics under appropriate conditions. At the same time, these effects arise in very different regimes, demonstrate very different dynamics, and are differently affected by the pulse imperfections. Understanding the connection between time crystals (in particular, infinite-temperature time crystals~\cite{khemaniNMR}) and long-lived coherences is an interesting and important, yet rather unexplored problem.

b) The presence of long-time tails along the $z$-axis which is perpendicular to the pulse driving field implies that the fundamental process for establishing long-time tails of the magnetization response may not be spin locking~\cite{slichter,abragam} as may appear in analogy with other similar effects~\cite{rhim,erofeev,suwelack}. 

c) Our simulations show that even if the direction of the driving during the $\pi$-pulses is chosen randomly for each spin, the long-lived coherences emerge and persist for long times, as long as the direction of the driving remains constant in time. This observation suggests that the origin of the long-lived coherences is primarily kinematic; it may arguably also involve many-body localization~\cite{abanin,yao}, transient prethermal regime~\cite{abanin2,roeck,rudner}, or spin-glass like behaviour.

d) The results shown in the present article were obtained for a fixed pulse imperfection $\epsilon=0.07$. Our numerical simulations with other values of $\epsilon$ demonstrate that qualitatively the same behavior occurs for other choices of $\epsilon\ll1$. The long-time spin coherences persist even when $\epsilon$ is chosen randomly for different spins, as long as it remains constant in time. 

e) We considered in this article spin systems of finite size, with the total number of spins $N_s\approx 20$. Our simulations demonstrate very modest quantitative changes as the system size has been varied from $N_s=10$ to $N_s=24$. Besides, our numerical results for $d=3$ agree with the reported experimental results for three-dimensional systems~\cite{li,dementyev,rovny,choichoi}.


Summarizing, we have investigated disordered dipolar-coupled quantum many-spin systems of different spatial dimensionality subjected to periodic driving (CPMG protocol). Depending on the dimensionality, such systems exhibit singularities in short-time spin dynamics, which are caused by statistical fluctuations in the dipolar spin-spin interaction strength. For all dimensionalities, we observed the long-lived spin polarization along the driving pulses, and along the axis conserved by the internal Hamiltonian ($z$-axis). The amplitude of the long-lived magnetization  depends on the inter-pulse time delay and on dimensionality. The Floquet operator analysis shows that the long-lived coherences $\mathcal{M}_x(t)$ contain comparable contributions from a large number of Floquet eigenstates. Our results imply that the long-lived coherences and subharmonic response are generic features of dipolar-coupled disordered spin systems, including two-, three-, and infinite-dimensional systems that are particularly relevant for practical applications. Developing specific protocols for such applications is an exciting avenue for further research.

\section*{Acknowledgements}
The authors would like to thank S. E. Barrett, A. Bleszynski-Jayich, D. Budker, N. de Leon, E. B. Fel'dman, B. V. Fine, G. E. Volovik, and N. Yao for stimulating discussions. This work was supported by the DARPA DRINQS program (contract D18AC00015KK1934). This work is part of the research programme NWO QuTech Physics Funding (QTECH, programme 172) with project number 16QTECH02, which is (partly) financed by the Dutch Research Council (NWO). The work was partially supported by the Kavli Institute of Nanoscience Delft.

\appendix

\section{Statistical analysis of disordered spin systems}
\label{app_StatAnalysis}

Let us discuss the effect of disorder in $d$ spatial dimensions on the Hahn echo  signal $\mathcal{M}_x^\text{H}(t)$ (denoted with the superscript ``H''). 
We are interested in the case of $T_2\gg T_2^*$, when the typical local fields $h_j$ are much larger than the typical dipolar interactions, such that the flip-flop term $S_{ix}S_{jx}+S_{iy}S_{jy}$ in dipolar Hamiltonian can be omitted~\cite{slichter,abragam,klauderanderson,dobrovitskiEns,feldman_lacelle}, see Appendix~\ref{app_flip}. In this case the $\pi$-pulse eliminates the inhomogeneous broadening 
${\cal H}_\text{L}$, and the system's response $\mathcal{M}_x(t)$ is determined only by the remaining Ising-like part of the dipolar Hamiltonian 
${\cal H}^0_\text{I}=\sum_{j>i} J_{ij} S_{iz}S_{jz}$, so the echo signal  
$\langle M_x^\text{H}(t)\rangle={\rm Tr}[M_x U_{0H}(t) M_x U^\dag_{0H}(t)]$, with 
$U_{0H}(t)=\exp{\left(-it{\cal H}^0_\text{I}\right)}$, can be directly calculated: 
\begin{equation}
\label{eq:mx0h}
\mathcal{M}_x^\text{H}(t)\propto\sum_i \prod_{j}\cos{\left(J_{ij}t/4\right)}. 
\end{equation}
For an ensemble where each spin $S_i$ has its own set of coupling constants $J_{ij}$, one should average Eq.~\eqref{eq:mx0h} over all possible positions of the spins $S_j$ in a $d$-dimensional sample of volume $V_d$. For dilute spins we can neglect the underlying crystal lattice, and replace summation over the lattice by integration over the whole space:
\begin{equation} 
\label{eqn_signal}
\mathcal{M}_x^\text{H}(t)=\left[\frac{1}{V_d}\int_{V_d}\cos{\left(b(\theta)t/r^3\right)} dv\right]^{N_s-1},
\end{equation}
where $dv$ is the volume element, and $b(\theta)=(1-3\cos^2\theta)/4$. 
To calculate the average for a macroscopically large spin ensemble, when $V_d\to\infty$ and $N_s\to\infty$ with a fixed spin density 
$f_s\equiv N_s/V_d$, the integral is re-written as~\cite{abragam,klauderanderson}
\begin{eqnarray}
\mathcal{M}_x^\text{H}(t)&=&\left[1-\frac{1}{V_d}\int_{V_d}\left(1-\cos{\left[b(\theta)t/r^3\right]}\right) dv\right]^{N_s-1}\\
&\approx&\exp{\left[-f_s\int_{V_d}\left(1-\cos{\left[b(\theta)t/r^3\right]}\right) dv\right]},
\end{eqnarray}
so that the integral in Eq.~\eqref{eqn_signal} is well defined at large $r$, where $\cos{\left[b(\theta)t/r^3\right]}\to 1$. 

Singularity of the spin dynamics is determined by the competition of two effects: the dipolar coupling constants $J_{ij}$ decrease with increasing $r$, but the number of spins $S_j$ which interact with the given spin $S_i$ grows with $r$ as the volume element $dv=r^{d-1} dr dA$ (where $dA$ is the surface element of the $(d-1)$-dimensional hypersphere of unit radius). For $d<6$, this growth is sufficiently slow, so the integration over $r$ can be extended to infinity, yielding the above-mentioned result (see Eq.~\eqref{eqn_hahn2} of the main text)
\begin{equation}
\label{eqn_hahn6}
\mathcal{M}_x^\text{H}(t)=\exp{\left(-\left|t/T_2\right|^{d/3}\right)}
\end{equation}
with the decay time 
$
T_2=\left[f_s\,\Lambda\,\int \left|b(\theta)\right|^{d/3} dA\right]^{-3/d}. 
$
The integral over the hypersphere is a numerical factor of order of one, and the quantity 
\begin{equation} 
\label{eqn_integral}
\Lambda=\frac{1}{3}\int_0^\infty \frac{1-\cos z}{z^{d/3+1}} dz=-\frac{1}{3}\cos\left(\frac{\pi d}{6}\right)\Gamma\left(-\frac{d}{3}\right),
\end{equation} 
which comes from integration over $r$, is also of order one, such that 
$T_2$ is determined just by the spin density $f_s$. By renormalizing the spin density, we can set $T_2=1$ as explained above.

For $d=2$, the singularity of the Hahn echo $\mathcal{M}_x^\text{H}(t)=\exp\left[-\left|t/T_2\right|^{2/3}\right]$ is strong: the initial echo decay is infinitely fast due to strong fluctuations in the positions of the spins at small distances $r_{ij}$. Although the typical distance between spins is of the order of $1/f_s^{1/d}=(V_d/N_s)^{1/d}$, a large fraction of spins has many neighbors at much smaller distances. Correspondingly, while the typical dipolar coupling is of the order of one (recall that we normalize $f_s$ to yield $T_2=1$), many realizations of  the disorder produce very large dipolar couplings $J_{ij}$. 

Of course, in real crystals, at extremely short times the initial decay rate is finite, because the distance between spins is limited by the crystal lattice constant, which in turn limits the maximal dipolar interaction strength. However, the corresponding times are extremely small, orders of magnitude smaller than $T_2^*$, and are irrelevant for the phenomena considered here.

For $d=3$, such fluctuations in $J_{ij}$ are less strong, and the singularity is weaker: $\mathcal{M}_x^\text{H}(t)=\exp\left(-\left|t/T_2\right|\right)$ has a cusp at $t=0$, but the initial rate of decay is finite. Still, for both $d=2$ and 3, the total spectral power of the resonance line is (formally) infinite. For $d=4$ and 5, the fluctuations are less pronounced, the total spectral power of the resonance is finite, and the singularity in $\mathcal{M}_x^\text{H}(t)$ is weak.

For $d\ge 6$, the integration over $r$ cannot be extended to infinity: the integral Eq.~\eqref{eqn_integral} diverges at small $z$ (which correspond to $r\to\infty$). This divergence means that the number of the spins $S_j$ coupled to the given spin $S_i$ grows too fast with increasing $r$. The contribution from the surface of the sample becomes important, dominating at larger $d$. The form of 
$\mathcal{M}^\text{H}_x(t)$ then depends on the sample shape and size, but in the limit $d\to\infty$ it again acquires a universal shape-independent Gaussian form. For any regular-shaped sample in the limit $d\to\infty$ all spins are located near the surface, and each spin pair is separated by almost the same distance, producing all-to-all interactions with a uniform coupling constant $J_{ij}\to J$. Eq.~\eqref{eqn_signal} yields $\mathcal{M}^\text{H}_x(t)=\cos^{N_s-1}(2Jt)$, and, since the value of $J$ scales as $1/\sqrt{N_s}$ for large $d$, in the limit $N_s\to\infty$ the Hahn echo signal is 
$\mathcal{M}_x^\text{H}(t)=\exp\left[-\left(t/T_2\right)^2\right]$, free of singularities, with finite spectral power of the resonance line. The effect of local fluctuations vanishes completely in this case, in stark contrast with $d=2$ and $d=3$ spin ensembles.

Since the key physical aspects of the spin dynamics are drastically different for the systems of different dimensionalities, it is reasonable to expect that the key features of the long-lived coherences would also differ drastically for different $d$. Surprisingly, our results have demonstrated this expectation to be incorrect.

\section{The role of the flip-flop processes} \label{app_flip}

\begin{figure}[]
  \centering
  \includegraphics[width=\columnwidth]{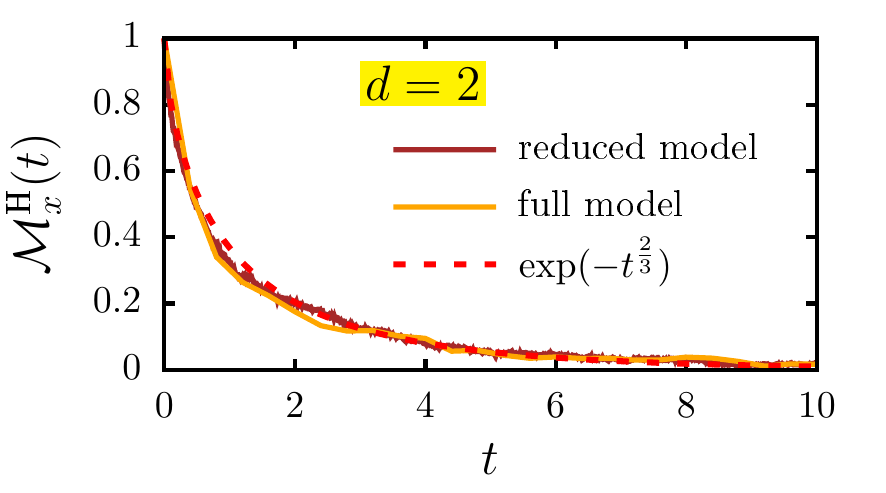}
  \caption{Hahn echo signal $\mathcal{M}^\text{H}_x(t)$ for two-dimensional disordered dipolar-coupled spin systems, calculated using the full and the reduced models, which differ by the presence of the flip-flop terms in the Hamiltonian. The results of the two models are in quantitative agreement with each other. The analytic approximation for Hahn echo in $d=2$ is $\mathcal{M}_x(t)=\exp(-t^{2/3})$. The system size is $N_s=18$, $T_2=1$, and $T_2^*\approx 0.02$.}
 \label{fig_2dredhhn}
\end{figure}

\begin{figure}[]
  \centering
  \includegraphics[width=\columnwidth]{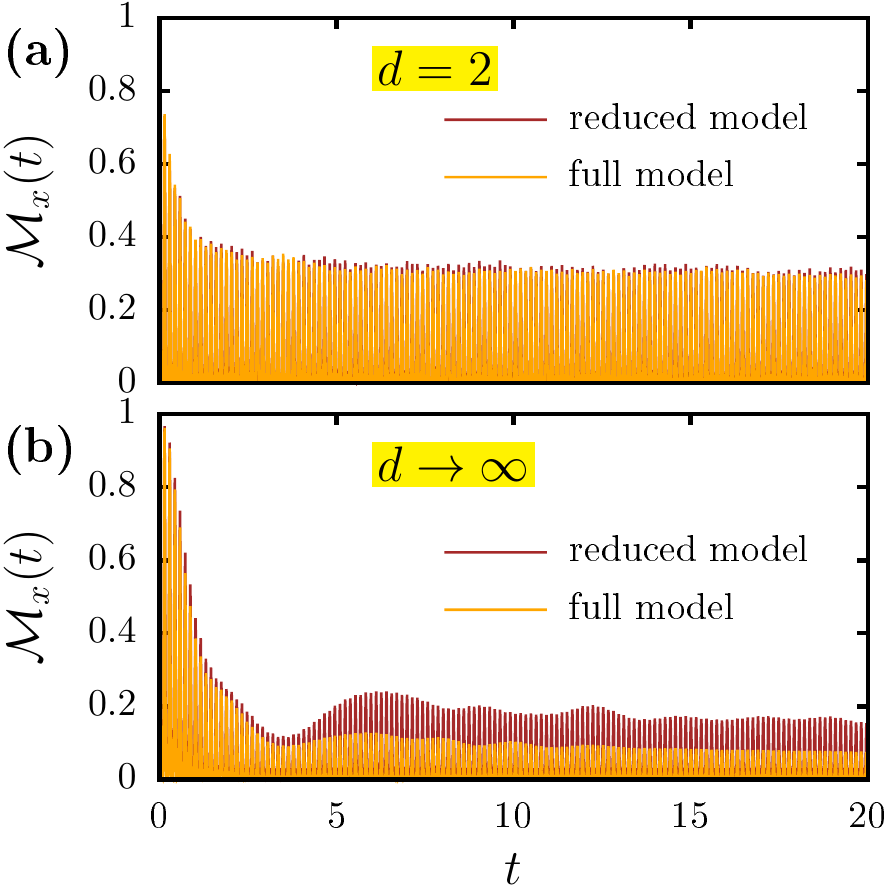}
  \caption{Multi-pulse CPMG signals, comparison between the reduced and the full models for $d=2$ \textbf{(a)}, and for $d\to\infty$ \textbf{(b)}. The results for the magnetization response $\mathcal{M}_x(t)$ are almost the same for 
  $d=2$, whereas for $d\to\infty$ the amplitudes of the long-lived coherences  differ by a factor of $\approx 2$. The results shown are for $N_s=20$ and short $\tau$; the other parameters are the same as in Fig.~\ref{fig_2dtime}.}
 \label{fig_ndredmdl}
\end{figure}

In this article, we consider a typical experimental situation where $T_2\gg T_2^*$, i.e.\ where the random local fields $h_j$ are much larger than the dipolar interactions  $J_{ij}$. Strong random local fields generally suppress the flip-flop terms 
$S_{ix}S_{jx}+S_{iy}S_{jy}=(1/2)\left[S_i^+ S_j^- + S_i^- S_j^+\right]$
in the dipolar Hamiltonian ${\cal H}_\text{I}$ given in Eq.~\eqref{eqn_hami}. It is reasonable to expect that the magnetization response does not change if the flip-flop terms are omitted, but the accumulation of the neglected terms at long times may become significant. Therefore, it is important to directly evaluate the role of the flip-flop terms. 

For the Hahn echo signal $\mathcal{M}^{\text{H}}_x(t)$ our simulations show that the impact of the flip-flop terms is insignificant: the curves remain practically the same, independent of whether we include the flip-flop terms or not, see Fig.~\ref{fig_2dredhhn}, where we refer to the Hamiltonian ${\cal H}^0_\text{I}=\sum_{j>i} J_{ij} S_{iz}S_{jz}$ without the flip-flop terms as the 
reduced
model and to the full dipolar Hamiltonian ${\cal H}_\text{I}$ as the 
full
model. 

The multi-pulse CPMG response $\mathcal{M}_x(t)$ for $d=2$ also demonstrates almost identical short-time behavior and long-lived coherences for the full and for the reduced models, as shown in Fig.~\ref{fig_ndredmdl}a. However, in the case $d\to\infty$ shown in Fig.~\ref{fig_ndredmdl}b, the calculated signals for both full and reduced models coincide only at short times, and exhibit quantitative difference later. Although both models clearly demonstrate long-lived coherences, the echo amplitude is approximately twice higher for the reduced model as compared with the full model. A possible reason for this discrepancy is in the geometry of the dipolar couplings. The flip-flop process between two spins is important if the dipolar coupling between the two spins is comparable to, or larger than, the difference in the respective local fields. In the situation considered here, with $T_2\gg T_2^*$, the above case can only occur if the local fields of two spins are accidentally similar. In $d=2$ systems, these two spins must be close to each other to ensure non-negligible dipolar coupling. In contrast, in $d\to\infty$ systems, these two spins can be at any distance to each other because the dipolar interaction is homogeneous, coupling each spin to all other spins. Thus, the probability for two spins to have accidentally similar local fields, and to  undergo a flip-flop process, is much larger in $d\to\infty$ systems than in $d=2$ systems. 


In this article we use the full model in our simulations, with the exception of the results for the single-pulse Hahn echo shown in Figs.~\ref{fig_2dtime} and \ref{fig_Idtime} of the main text.

\section{The role of pulse errors} \label{app_errors}

\begin{figure}[b]
  \centering
  \includegraphics[width=\columnwidth]{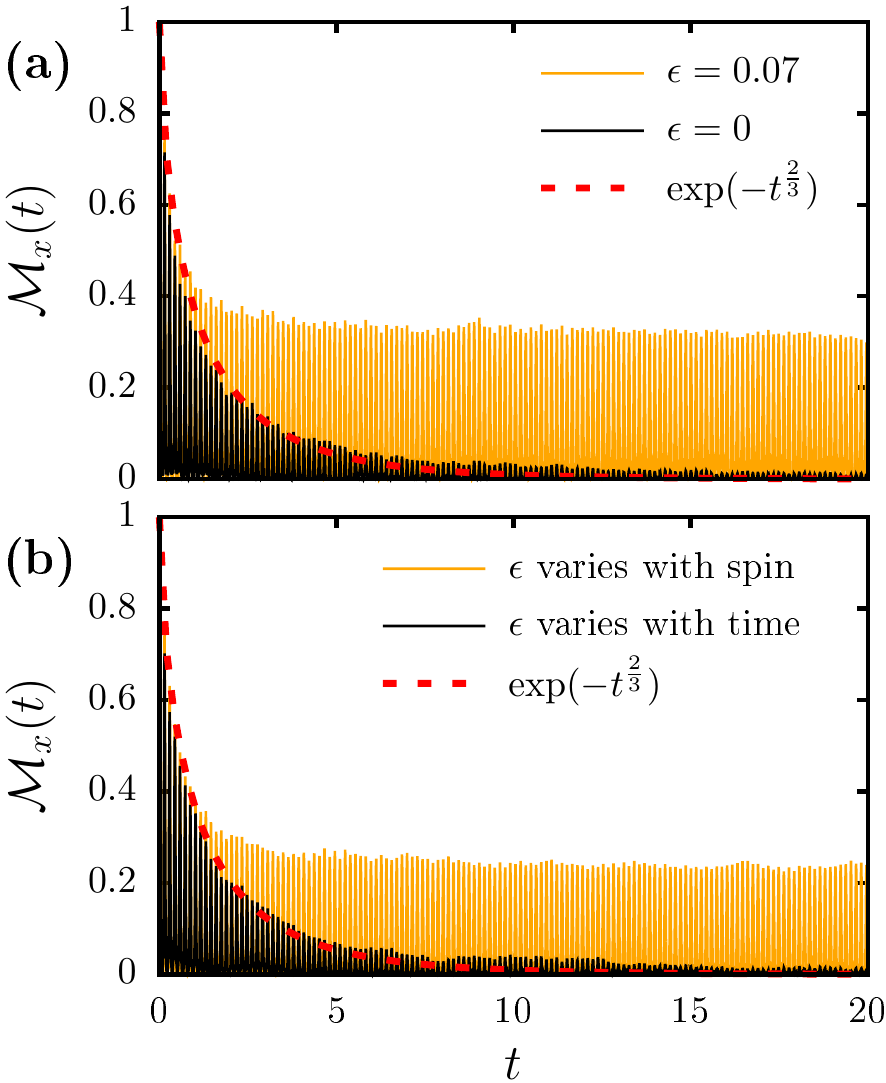}
  \caption{Multi-pulse CPMG signals for a two-dimensional ($d=2$) disordered dipolar-coupled spin systems, demonstrating the role of the pulse errors in appearance of the long-lived coherences. 
  \textbf{(a)}. Comparison between the decay of coherence $\mathcal{M}_x(t)$ for perfect pulses with $\epsilon=0$ (black solid line) and for imperfect pulses with $\epsilon=0.07$ (orange solid line). In the case of perfect pulses, the decay of the CPMG signal is fast, and follows the decay of the Hahn echo (red dashed line, showing the theoretically predicted curve $\exp{[-t^{2/3}]}$), while for the imperfect pulses a long-lived coherence is clearly seen.
  \textbf{(b)}. Comparison between the decay of coherence $\mathcal{M}_x(t)$ for imperfect pulses. Black solid line represents the situation when the pulse error (the value of $\epsilon$) is the same for all spins but randomly varies in time, being sampled uniformly from the interval $[-0.07,0.07]$ for each new pulse. Orange solid line corresponds to the case when the pulse errors stay constant in time but vary in a random way from one spin to another (sampled uniformly from the interval $[-0.07,0.07]$). In the case of the time-varying pulse errors, the decay of the CPMG signal is fast, and follows the decay of the Hahn echo (red dashed line, showing the theoretically predicted curve $\exp{[-t^{2/3}]}$). However, when the pulse errors remain constant in time, a long-lived coherence is clearly seen, even though the pulse errors have different values for different spins.
  For both panels, the results were obtained for $N_s=18$ and averaged over 200 samples; all other parameters are the same as in Fig.~\ref{fig_2dtime}.}
 \label{fig_2dextra}
\end{figure}

In all simulations above, we considered the case where the pulse error $\epsilon$ is constant in time, and is the same for all spins. Fig.~\ref{fig_2dextra} demonstrates the role of pulse errors in producing the long-lived coherences. In panel (a) we show that ideal pulses, with $\epsilon=0$, do not produce the long-lived coherences: the decay of the multi-pulse CPMG signal $\mathcal{M}_x(t)$ is fast, and follows the decay of the single-pulse Hahn echo; the latter is in perfect agreement with the theoretically predicted form $\mathcal{M}_x^\text{H}(t)=\exp{\left[-(t/T_2)^{2/3}\right]}$. 
The absence of the long-lived coherences for ideal pulses is a straightforward  consequence of the fact that the dipolar coupling Hamiltonian ${\cal H}_\text{I}=\sum_{i>j} (J_{ij}/2) \left [2S_{iz}S_{jz}-S_{ix}S_{jx}-S_{iy}S_{jy}\right]$ is invariant with respect to ideal $\pi$-pulses along the $x$-axis (i.e.\ with respect to simultaneous change $S_{jz} \to -S_{jz}$ and $S_{jy} \to -S_{jy}$ for all spins). 

The panel (b) shows that the pulse errors can be different for different spins, and still produce long-lived coherences, as long as the pulse errors remain  constant in time. The orange solid line shows the multi-pulse CPMG signal $\mathcal{M}_x(t)$ for the case when the pulse error $\epsilon$ is chosen randomly for each spin (sampled uniformly from the interval $[-0.07,0.07]$), but is kept constant in time. In contrast, even if the pulse error is the same for all spins, but varies in time from one pulse to the next (e.g.\ sampled uniformly from the interval $[-0.07,0.07]$ for each pulse anew), the ensemble coherence in CPMG experiment exhibits fast decay (black line), coinciding with the decay of the single-pulse Hahn echo (red dashed line). The results shown in Fig.~\ref{fig_2dextra} have been obtained for $N_s=18$; all other parameters are the same as in Fig.~\ref{fig_2dtime}.

\section{Weighted contribution of different Floquet states to the long-lived coherences}

In Sec.~\ref{sec:Floquet} we presented detailed analysis of the Floquet states giving rise to the long-lived coherences. We carefully distinguished and studied separately the two contributions to the CPMG signal $\mathcal{M}_x(t)$ (see Eq.~\ref{eqn_px}): the values of the matrix elements $|\langle\psi_j|M_x|\psi_k\rangle|^2$ between different pairs of the Floquet states $|\psi_j\rangle$ and $|\psi_k\rangle$, see Fig.~\ref{fig_2d} for $d=2$ and Fig.~\ref{fig_5d} for $d=5$, and the number $P(\Delta\phi)$ of contributing pairs of states, see Fig.~\ref{fig_specdiff}.

Yet another way to quantify the contribution of different Floquet states to the long-lived coherences is to consider the combined effect of both contributions, and study the weighted distribution $\Sigma(\Delta\phi)$ of the quasienergy differences $\phi_j-\phi_k$, formally defined as
\begin{eqnarray} 
\nonumber
\Sigma(\Delta\phi)&=&\sum_{j,k} |\langle\psi_j|M_x|\psi_k\rangle|^2 \ \Bigl[\delta(|\phi_j-\phi_k|-\Delta\phi)\\
&+&\delta(2\pi-|\phi_j-\phi_k|-\Delta\phi)\Bigr],
\label{eq:sigma}
\end{eqnarray}
such that each pair of the Floquet states having the given value of $\Delta\phi$ enters the sum $\Sigma(\Delta\phi)$ with the weight coefficient equal to $|\langle\psi_j|M_x|\psi_k\rangle|^2$; this corresponds to summation along diagonals of Figs.~\ref{fig_2d} and \ref{fig_5d}. Two $\delta$-functions in the formula reflect the fact that the quantity $\Delta\phi$ is downfolded to the interval $[0,\pi]$ , i.e.\ we take
$\Delta\phi=|\phi_j-\phi_k|$ when $|\phi_j-\phi_k|\le\pi$ and $\Delta\phi=2\pi-|\phi_j-\phi_k|$ when $|\phi_j-\phi_k|>\pi$; this downfolding reflects the symmetries of the summands in Eq.~\ref{eqn_px}. Note that the quantity $\Sigma(\Delta\phi)$ is proportional to the cosine Fourier transform of the CPMG signal $\mathcal{M}_x(2\tau m)$.

\begin{figure}
  \centering
  \includegraphics[width=\columnwidth]{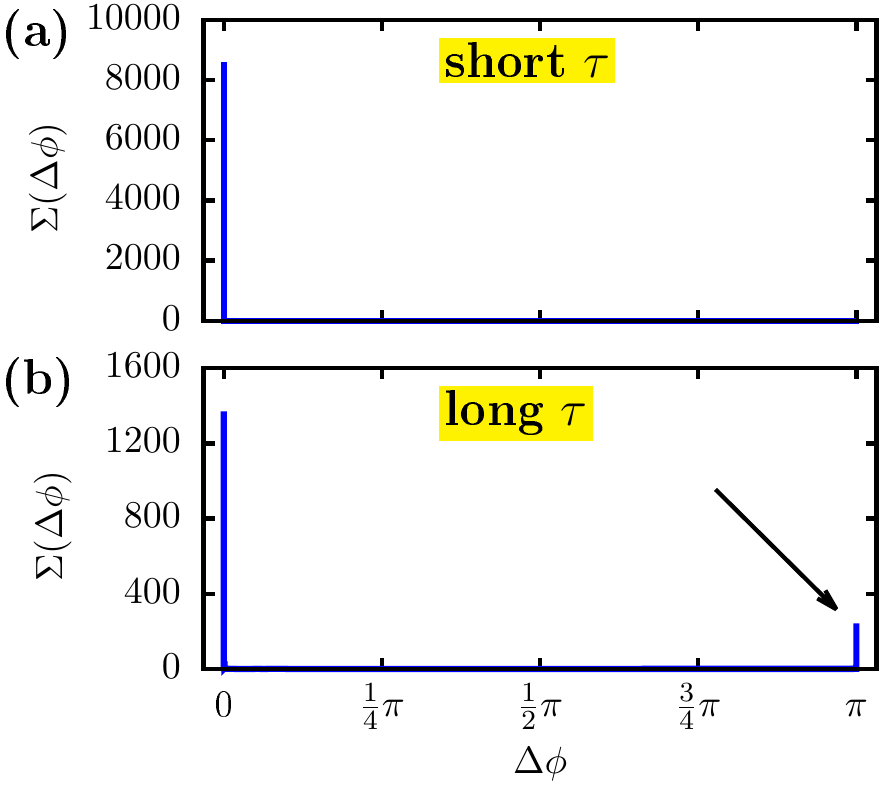}
  \caption{Weighted distribution $\Sigma(\Delta\phi)$ of the quasienergy differences $|\phi_j-\phi_k|$ for a $d=2$ disordered spin system with $N_s=14$, at short $\tau$ \textbf{(a)} and long $\tau$ \textbf{(b)}. The sharp peaks at $\Delta\phi=0$ and at $\Delta\phi=\pi$ correspond to the long-lived coherences and to the subharmonic response, respectively. The distribution includes averaging over many realizations of the disorder. The bin width is $2\beta'=\pi\cdot 10^{-5}$, i.e.\ for each value of $\Delta\phi$ the sum includes all Floquet states with  $|\phi_j-\phi_k|\in\left[\Delta\phi-\beta',\Delta\phi+\beta'\right]$ or
  $2\pi-|\phi_j-\phi_k|\in\left[\Delta\phi-\beta',\Delta\phi+\beta'\right]$. To avoid double counting, only the states with $\phi_j\ge\phi_k$ are included in $\Sigma(\Delta\phi)$. The normalization is chosen such that $\int\Sigma(\Delta\phi)d(\Delta\phi)=1$. All other simulation parameters are the same as in Fig.~\ref{fig_2dtime}.}
 \label{fig_msx_comp}
\end{figure}

The graphs of $\Sigma(\Delta\phi)$ are presented in Fig.~\ref{fig_msx_comp}, where instead of an infinitely narrow delta function we used bins of finite width $2\beta'=\pi\cdot 10^{-5}$, i.e.\ the sum in Eq.~\ref{eq:sigma} includes  all states whose quasienergies satisfy the condition  $|\phi_j-\phi_k|\in\left[\Delta\phi-\beta',\Delta\phi+\beta'\right]$ or 
$2\pi-|\phi_j-\phi_k|\in\left[\Delta\phi-\beta',\Delta\phi+\beta'\right]$. 
The distinctive peak at $|\phi_j-\phi_k|\approx\pi$, which corresponds to emergence of the subharmonic response, appears only for long $\tau$. The peak 
at $\phi_j-\phi_k\approx 0$, which corresponds to the long-lived response, is formed by the pairs of eigenstates having almost the same quasienergy, and the main contribution to this peak comes from the situation $j=k$, when $|\psi_j\rangle$ and $|\psi_k\rangle$ are the same.

\end{document}